\PassOptionsToPackage{unicode}{hyperref}
\PassOptionsToPackage{hyphens}{url}
\documentclass[
  english,
  onecolumn]{article}
\usepackage{xcolor}
\usepackage[margin=2cm,headheight=15pt]{geometry}
\usepackage{amsmath,amssymb}
\usepackage{iftex}
  \usepackage[T1]{fontenc}
  \usepackage[utf8]{inputenc}
  \usepackage{textcomp} 
\usepackage{lmodern}
\IfFileExists{upquote.sty}{\usepackage{upquote}}{}
\IfFileExists{microtype.sty}{
  \usepackage[]{microtype}
  \UseMicrotypeSet[protrusion]{basicmath} 
}{}
\makeatletter
\@ifundefined{KOMAClassName}{
  \IfFileExists{parskip.sty}{%
    \usepackage{parskip}
  }{
    \setlength{\parindent}{0pt}
    \setlength{\parskip}{6pt plus 2pt minus 1pt}}
}{
  \KOMAoptions{parskip=half}}
\makeatother
\usepackage[bidi=default,shorthands=off,]{babel}
\providecommand{\tightlist}{%
  \setlength{\itemsep}{0pt}\setlength{\parskip}{0pt}}
\usepackage[numbers,sort,compress]{natbib}
\usepackage{fancyhdr}
\usepackage{titlesec}
\usepackage{bookmark}
\IfFileExists{xurl.sty}{\usepackage{xurl}}{} 
\hypersetup{
  pdftitle={Formal Verification of a Token Sale Launchpad: A Compositional Approach in Dafny},
  pdfauthor={Evgeny Ukhanov; Aurora Labs},
  pdflang={en},
  hidelinks,
  pdfcreator={LaTeX via pandoc}}

\title{Formal Verification of a Token Sale Launchpad: A Compositional
Approach in Dafny}
\author{Evgeny Ukhanov \and Aurora Labs}
\date{August 2025}

\begin{document}
\maketitle

\section{Abstract}\label{abstract}

The proliferation of decentralized financial (DeFi) systems and smart
contracts has underscored the critical need for software correctness.
Bugs in such systems can lead to catastrophic financial losses. Formal
verification offers a path to achieving mathematical certainty about
software behavior \citep{borkowski2006formal}. This paper presents the
formal verification of the core logic for a token sale launchpad,
implemented and proven correct using the Dafny programming language and
verification system. We detail a compositional, bottom-up verification
strategy, beginning with the proof of fundamental non-linear integer
arithmetic properties, and building upon them to verify complex business
logic, including asset conversion, time-based discounts, and capped-sale
refund mechanics. The principal contributions are the formal proofs of
critical safety and lifecycle properties. Most notably, we prove that
\textbf{refunds in a capped sale can never exceed the user's original
deposit amount}, and that the precision loss in round-trip financial
calculations is strictly bounded. Furthermore, we verify the complete
lifecycle logic, including user withdrawals under various sale mechanics
and the correctness of post-sale token allocation, vesting, and
claiming. This work serves as a comprehensive case study in applying
rigorous verification techniques to build high-assurance financial
software \citep{atzei2017survey}.

\section{1. Introduction}\label{introduction}

The domain of financial software, particularly in the context of
blockchain and smart contracts, operates under a unique and unforgiving
paradigm: deployed code is often immutable, and flaws can be exploited
for immediate and irreversible financial gain
\citep{atzei2017survey, weir2018formal}. Traditional software testing,
while essential, is inherently incomplete as it can only validate a
finite set of execution paths. Formal verification addresses this
limitation by using mathematical logic to prove properties about a
program's behavior across \emph{all} possible inputs that satisfy its
preconditions \citep{woodcock2009formal, clarke1996formal}.

This paper focuses on the formal verification of a token sale launchpad
contract. The core challenge lies in reasoning about complex
interactions between multiple components: price-based asset conversions,
application of percentage-based bonuses (discounts), and state
transitions governed by sale mechanics (e.g., fixed-price vs.~price
discovery), all while handling the subtleties of integer arithmetic.

Our tool of choice is \textbf{Dafny}, a verification-aware programming
language that integrates specification and implementation
\citep{leino2010dafny}. Dafny allows programmers to annotate their code
with formal contracts, such as preconditions (\texttt{requires}),
postconditions (\texttt{ensures}), and loop invariants
(\texttt{invariant}). These annotations, along with the program code,
are translated by the Dafny verifier into logical formulas, which are
then dispatched to an automated Satisfiability Modulo Theories (SMT)
solver, typically Z3 \citep{demoura2008z3}. If the solver can prove all
formulas, the program is deemed correct with respect to its
specification.

The primary objective of this work is to construct a fully verified
model of the launchpad's core logic. We demonstrate how a carefully
layered architecture enables the verification of a complex system by
decomposing the proof effort into manageable, reusable components. We
will present the key modules, the mathematical properties they
guarantee, and the overarching safety lemmas that emerge from their
composition.

\section{2. System Architecture and Verification
Strategy}\label{system-architecture-and-verification-strategy}

The verification effort is structured around a \textbf{compositional,
bottom-up approach}, which is crucial for managing complexity
\citep{almeida2007compositional, chen2021compositional}. The system is
decomposed into a hierarchy of modules, where each higher-level module
relies on the proven correctness of the modules below it. This isolates
reasoning and makes the overall verification problem tractable.

The architecture consists of the following layers:

\begin{enumerate}
\def\labelenumi{\arabic{enumi}.}
\tightlist
\item
  \textbf{\texttt{MathLemmas}:} The foundational layer. It provides
  proofs for fundamental, non-trivial properties of non-linear integer
  arithmetic (multiplication and division), which are not natively
  understood by SMT solvers.
\item
  \textbf{\texttt{AssetCalculations} \& \texttt{Discounts}:} The
  business logic primitives layer. These modules define the core
  financial calculations (asset conversion, discount application) and
  use lemmas from \texttt{MathLemmas} to prove their essential
  properties, such as monotonicity and round-trip safety.
\item
  \textbf{\texttt{Config}, \texttt{Investments}:} The data modeling
  layer. These modules define the primary data structures of the system,
  including the main \texttt{Config} datatype which encapsulates all
  sale parameters and rules.
\item
  \textbf{\texttt{Deposit}, \texttt{Withdraw}, \texttt{Claim},
  \texttt{Distribution}:} The workflow specification layer. These
  modules compose primitives from lower layers to define the complete,
  pure specifications for all user and administrative interactions,
  including deposits with refund logic, withdrawals, post-sale token
  claims with vesting, and stakeholder distributions.
\item
  \textbf{\texttt{Launchpad}:} The top-level state machine. This module
  defines the complete contract state and models all lifecycle
  transitions by orchestrating the verified workflows from the layer
  below.
\end{enumerate}

A key pattern employed throughout the codebase is the
\textbf{Specification-Implementation Separation}
\citep{leino2017tutorial}. For most critical operations, a
\texttt{function} ending in \texttt{...Spec} defines the pure
mathematical contract. This allows us to reason about the system's logic
at an abstract, mathematical level.

\section{3. Foundational Layer: Verification of Non-Linear Integer
Arithmetic}\label{foundational-layer-verification-of-non-linear-integer-arithmetic}

Reasoning about the multiplication and division of integers is a
well-known challenge in automated verification
\citep{monniaux2008pitfalls, audemard2021certified}. SMT solvers are
highly effective for linear integer arithmetic, but non-linear
properties often require explicit proof guidance. The
\texttt{MathLemmas} module provides this guidance by establishing a set
of trusted, reusable lemmas for the rest of the system.

The core of financial calculations in this system is scaling a value
\(x\) by a rational factor \(y/k\), implemented using integer arithmetic
as \((x \cdot y) / k\). The following key lemmas were proven from first
principles:

\begin{itemize}
\item
  \textbf{Monotonicity with Favorable Scaling
  (\texttt{Lemma\_MulDivGreater\_From\_Scratch}):} This lemma proves
  that if the scaling factor is greater than or equal to 1, the result
  is no less than the original amount.

  \begin{itemize}
  \tightlist
  \item
    \textbf{Property 1.} \(\forall x, y, k \in \mathbb{N}\) where
    \(x > 0\), \(k > 0\), and \(y \ge k\) \[\frac{x \cdot y}{k} \ge x\]
    This is crucial for proving that conversions at a stable or
    favorable price do not result in a loss of principal.
  \end{itemize}
\item
  \textbf{Strict Monotonicity with Highly Favorable Scaling
  (\texttt{Lemma\_MulDivStrictlyGreater\_From\_Scratch}):} Due to
  integer division truncation, \(y > k\) is insufficient to guarantee
  \((x \cdot y) / k > x\). This lemma establishes a stronger
  precondition to ensure a strict increase.

  \begin{itemize}
  \tightlist
  \item
    \textbf{Property 2.} \(\forall x, y, k \in \mathbb{N}\) where
    \(x > 0\), \(k > 0\), and \(y \ge 2k\):\\
    \[\frac{x \cdot y}{k} > x\] This is used to prove that a
    significantly favorable price or a large bonus yields a tangible
    gain.
  \end{itemize}
\item
  \textbf{Round-trip Truncation and Bounded Loss
  (\texttt{Lemma\_DivMul\_Bounds}):} This lemma formalizes the
  fundamental property of integer division: information may be lost, but
  this loss is strictly bounded.

  \begin{itemize}
  \tightlist
  \item
    \textbf{Property 3.} \(\forall x, y \in \mathbb{N}\) where
    \(y > 0\):
    \[ \left( \frac{x}{y} \right) \cdot y \le x \quad \land \quad x - \left( \frac{x}{y} \right) \cdot y < y \]
    This property is the cornerstone for proving the safety of
    round-trip calculations. It not only guarantees that a reverse
    operation cannot create value, but also establishes a precise upper
    bound on the precision loss, which cannot exceed the value of the
    divisor \texttt{y}.
  \end{itemize}
\end{itemize}

These foundational lemmas abstract away the complexities of integer
arithmetic, allowing higher-level modules to reason about calculations
in terms of simple inequalities.

\section{4. Core Business Logic
Verification}\label{core-business-logic-verification}

Building upon the \texttt{MathLemmas} foundation, we verify the core
business logic components.

\subsection{\texorpdfstring{4.1. Asset Conversion
(\texttt{AssetCalculations})}{4.1. Asset Conversion (AssetCalculations)}}\label{asset-conversion-assetcalculations}

This module defines the logic for converting a base \texttt{amount} into
assets based on a price fraction \texttt{saleToken\ /\ depositToken}.
The specification is:
\texttt{CalculateAssetsSpec(amount,\ dT,\ sT)\ :=\ (amount\ *\ sT)\ /\ dT}

The module provides lemmas that instantiate the generic mathematical
properties for this specific context. For instance,
\texttt{Lemma\_CalculateAssets\_IsGreaterOrEqual} proves that
\texttt{CalculateAssetsSpec(...)\ \textgreater{}=\ amount} if
\texttt{sT\ \textgreater{}=\ dT}, by directly invoking
\texttt{Lemma\_MulDivGreater\_From\_Scratch}.

A critical property for refund safety is the \textbf{round-trip
inequality with bounded loss}, proven in
\texttt{Lemma\_AssetsRevert\_RoundTrip\_bounds}
\citep{bhargavan2016formal}. It states that converting an amount to
assets and then back cannot result in a gain, and furthermore, that any
loss due to truncation is strictly bounded.

\begin{itemize}
\tightlist
\item
  \textbf{Property 4 (Asset Conversion Round-Trip Safety and Bounded
  Loss).} Let \(Assets(w) := \text{CalculateAssetsSpec}(w, dT, sT)\) and
  \(Revert(a) := \text{CalculateAssetsRevertSpec}(a, dT, sT)\). Then for
  \(w > 0\):
  \[ \text{Assets}(w) > 0 \implies \text{Revert}(\text{Assets}(w)) \le w \]
  Moreover, the lemma proves a stronger property: the scaled difference
  between the original and reverted amounts will never exceed the sum of
  the price fraction's terms:
  \[ (w - \text{Revert}(\text{Assets}(w))) \cdot sT < dT + sT \] This
  guarantee is crucial as it proves that financial loss from rounding
  errors is predictable and has a hard ceiling.
\end{itemize}

\subsection{\texorpdfstring{4.2. Time-Based Discounts
(\texttt{Discounts})}{4.2. Time-Based Discounts (Discounts)}}\label{time-based-discounts-discounts}

This module implements percentage-based bonuses. It uses fixed-point
arithmetic with a \texttt{MULTIPLIER} of 10000 to represent percentages
with four decimal places. It also verifies a critical business rule:
discount periods must not overlap.

\begin{itemize}
\tightlist
\item
  \textbf{Property 5 (Discount Non-Overlap).} For a sequence of
  discounts \(D\), the following predicate holds:
  \[ \forall i, j. (0 \le i < j < |D|) \implies (D_i.\text{endDate} \le D_j.\text{startDate} \lor D_j.\text{endDate} \le D_i.\text{startDate}) \]
\end{itemize}

Dafny successfully proves that this property implies the uniqueness of
any active discount at a given time (
\texttt{Lemma\_UniqueActiveDiscount}), which is essential for ensuring
that deposit calculations are deterministic and unambiguous
\citep{grishchenko2018semantic}. Similar to asset conversions, the
module also proves the round-trip safety for applying and reverting a
discount (\texttt{Lemma\_WeightOriginal\_RoundTrip\_lte}).

\section{5. Top-Level Specification and State Machine
Verification}\label{top-level-specification-and-state-machine-verification}

The verified components are composed at the top layers to model the
complete system behavior.

\subsection{\texorpdfstring{5.1. The Deposit Workflow and Refund Safety
(\texttt{Deposit}
module)}{5.1. The Deposit Workflow and Refund Safety (Deposit module)}}\label{the-deposit-workflow-and-refund-safety-deposit-module}

This module specifies the end-to-end logic for a user deposit. The main
function, \texttt{DepositSpec}, branches based on the sale mechanic. The
most complex case is \texttt{DepositFixedPriceSpec}, which handles
deposits into a sale with a hard cap ( \texttt{saleAmount}). If a
deposit would cause the total sold tokens to exceed this cap, a partial
refund must be calculated.

The paramount safety property for this entire system is ensuring that
this refund never exceeds the user's initial deposit. This is formally
stated and proven in \texttt{Lemma\_RefundIsSafe}.

\begin{itemize}
\tightlist
\item
  \textbf{Property 6 (Ultimate Refund Safety).} For a valid
  configuration and any deposit \texttt{amount}, the calculated
  \texttt{refund} adheres to the following inequality:
  \[ \text{refund} \le \text{amount} \]
\end{itemize}

The proof of this high-level property is a testament to the
compositional strategy. It is not proven from first principles but by
orchestrating a chain of previously-proven, and now stronger, lemmas:

\begin{enumerate}
\def\labelenumi{\arabic{enumi}.}
\tightlist
\item
  \texttt{Lemma\_CalculateAssetsRevertSpec\_Monotonic} is used to show
  that the reverted value of the \emph{excess} assets is less than or
  equal to the reverted value of the \emph{total} assets.
\item
  \texttt{Lemma\_AssetsRevert\_RoundTrip\_bounds} proves not only that
  the reverted value of the total assets is less than or equal to the
  user's initial weighted amount (\texttt{\textless{}=\ w}), but also
  that any precision loss from this round-trip conversion is strictly
  bounded.
\item
  \texttt{Lemma\_CalculateOriginalAmountSpec\_Monotonic} shows that
  reverting the discount on a smaller amount yields a smaller result.
\item
  \texttt{Lemma\_WeightOriginal\_RoundTrip\_bounds} proves that the
  final original amount is less than or equal to the user's initial
  deposit amount, and more powerfully, that this round-trip operation
  can result in a loss of at most one minimal unit.
\end{enumerate}

By chaining these proven inequalities, Dafny confirms the ultimate
safety property: \texttt{refund\ \textless{}=\ amount}. This guarantee
is built upon a foundation of lemmas that provide much stronger,
explicit bounds on precision loss at each stage. The proof relies on the
facts that precision loss from asset conversion is strictly bounded by
the terms of the price fraction (as proven in
\texttt{Lemma\_AssetsRevert\_RoundTrip\_bounds}), and that the loss from
applying and reverting a discount is at most one minimal unit (from
\texttt{Lemma\_WeightOriginal\_RoundTrip\_bounds}). The final proof
confirms that the cumulative effect of these individually-bounded
truncations can never compound in a way that would violate the top-level
safety property, thus providing a mathematical guarantee against a
critical class of financial bugs.

\subsection{\texorpdfstring{5.2. The Withdrawal Workflow
(\texttt{Withdraw}
module)}{5.2. The Withdrawal Workflow (Withdraw module)}}\label{the-withdrawal-workflow-withdraw-module}

The \texttt{Withdraw} module provides the formal specification for users
to retrieve their funds under specific circumstances, such as a failed
sale or during the \texttt{PriceDiscovery} phase. The logic is
bifurcated based on the sale mechanic:

\begin{itemize}
\tightlist
\item
  \textbf{Fixed-Price Withdrawals (\texttt{WithdrawFixedPriceSpec}):} In
  a failed sale, this models an ``all-or-nothing'' withdrawal. The
  user's entire investment is returned, and their corresponding weight
  is removed from the \texttt{totalSoldTokens}.
\item
  \textbf{Price-Discovery Withdrawals
  (\texttt{WithdrawPriceDiscoverySpec}):} This models a partial or full
  withdrawal where the user's contribution is re-evaluated. The
  specification guarantees that \texttt{totalSoldTokens} is correctly
  reduced by the precise difference in the user's weight, ensuring the
  integrity of the final price calculation.
\end{itemize}

The verification of this module ensures that state changes related to
withdrawals are handled safely, preventing accounting errors.

\subsection{\texorpdfstring{5.3. Token Claim and Vesting Logic
(\texttt{Claim}
module)}{5.3. Token Claim and Vesting Logic (Claim module)}}\label{token-claim-and-vesting-logic-claim-module}

The \texttt{Claim} module formalizes the post-sale logic for users to
claim their purchased tokens. Its verification provides mathematical
guarantees about the correctness of token allocation and vesting
schedules. Key verified components include:

\begin{itemize}
\tightlist
\item
  \textbf{User Allocation (\texttt{UserAllocationSpec}):} This function
  specifies the total tokens a user is entitled to based on their final
  \texttt{weight} and the total \texttt{totalSoldTokens}. The lemma
  \texttt{Lemma\_UserAllocationSpec} proves its core mathematical
  properties, ensuring allocations are fair and predictable.
\item
  \textbf{Vesting Calculation (\texttt{CalculateVestingSpec}):} This
  function models a standard vesting schedule with a cliff and linear
  release. The core safety property, proven in
  \texttt{Lemma\_CalculateVestingSpec\_Monotonic}, guarantees that the
  vested amount never decreases as time moves forward.
\item
  \textbf{Available to Claim (\texttt{AvailableForClaimSpec}):} This
  function composes the allocation and vesting logic to determine the
  exact amount a user can claim at a specific time.
\end{itemize}

The verification of this module is critical for ensuring that the final
distribution of tokens strictly adheres to the sale's rules and vesting
commitments.

\subsection{\texorpdfstring{5.4. Post-Sale Distribution
(\texttt{Distribution}
module)}{5.4. Post-Sale Distribution (Distribution module)}}\label{post-sale-distribution-distribution-module}

The \texttt{Distribution} module specifies the administrative function
of distributing tokens to project stakeholders (e.g., the team,
partners, and the solver) after a successful sale. The core function,
\texttt{GetFilteredDistributionsSpec}, formally defines the logic for
identifying which stakeholders are eligible for the next distribution
round by filtering out those who have already received their tokens. The
verification ensures this process is deterministic and complete,
preventing accounts from being either skipped or paid multiple times.

\subsection{\texorpdfstring{5.5. The Contract State Machine
(\texttt{Launchpad}
module)}{5.5. The Contract State Machine (Launchpad module)}}\label{the-contract-state-machine-launchpad-module}

The \texttt{Launchpad} module represents the apex of the verification
hierarchy. It defines the global state of the contract within the
immutable \texttt{AuroraLaunchpadContract} datatype and models all of
its lifecycle transitions \citep{hirai2017defining}. The state
representation is comprehensive, encapsulating not only the core
financial tallies but also tracking the lifecycle of post-sale events,
such as stakeholder distributions (\texttt{distributedAccounts}) and
individual vesting claims (\texttt{individualVestingClaimed}).

The \texttt{GetStatus} function provides a pure, verifiable definition
of the contract's status (e.g., \texttt{NotStarted}, \texttt{Ongoing},
\texttt{Success}), which serves as the basis for enforcing
state-dependent business rules. This module includes critical lemmas
that prove the logical integrity of the state machine itself:

\begin{itemize}
\tightlist
\item
  \textbf{Mutual Exclusion (\texttt{Lemma\_StatusIsMutuallyExclusive}):}
  The contract cannot be in two different states simultaneously.
\item
  \textbf{Temporal Progression
  (\texttt{Lemma\_StatusTimeMovesForward}):} The contract progresses
  logically through its lifecycle as time advances.
\item
  \textbf{Terminal States
  (\texttt{Lemma\_StatusFinalStatesAreTerminal}):} Once a final state
  (\texttt{Success}, \texttt{Failed}, \texttt{Locked}) is reached, it
  cannot be exited.
\end{itemize}

The core of this module is the set of functions that model the
contract's state transitions. Each transition is a pure function that
takes the previous state \(\Sigma\) and returns a new state \(\Sigma'\),
thereby providing a complete and auditable specification of the
contract's behavior. Key verified transitions include:

\begin{itemize}
\tightlist
\item
  \textbf{\texttt{DepositSpec}:} Models the full state transition upon a
  user deposit. It enforces that deposits are only possible during the
  \texttt{Ongoing} state and delegates all complex financial logic
  (including refund calculations) to the pre-verified
  \texttt{Deposit.DepositSpec} function.
\item
  \textbf{\texttt{WithdrawSpec}:} Specifies the logic for users to
  withdraw their funds. Its preconditions ensure this action is only
  permissible in valid states (e.g., \texttt{Failed}, \texttt{Locked},
  or during an \texttt{Ongoing} price discovery sale). The function
  orchestrates the state change by invoking the verified
  \texttt{Withdraw} module.
\item
  \textbf{\texttt{ClaimSpec} and \texttt{ClaimIndividualVestingSpec}:}
  These functions model the post-sale token claim process for public
  participants and private stakeholders, respectively. They enforce that
  claims can only occur after a \texttt{Success} state is reached and
  correctly update the user's \texttt{claimed} balance by delegating the
  complex allocation and vesting calculations to the \texttt{Claim}
  module.
\item
  \textbf{\texttt{DistributeTokensSpec}:} Defines the administrative
  state transition for distributing tokens to project stakeholders. This
  action is guarded to ensure it only happens post-success and
  orchestrates the update of the \texttt{distributedAccounts} list by
  calling the verified \texttt{Distribution} module.
\end{itemize}

The verification of these top-level transitions is a powerful
demonstration of the compositional strategy. The proofs at this layer do
not re-verify the complex financial safety properties (like refund
safety or vesting curve monotonicity). Instead, they focus solely on
proving that the global state fields are updated correctly based on the
outputs from the already-proven workflow functions. This separation of
concerns reduces the safety of the entire system to the correctness of
its orchestration logic, given the proven correctness of its parts
\citep{cohen2017certified}.

\section{6. Limitations and Future
Work}\label{limitations-and-future-work}

While the compositional verification in Dafny provides a high degree of
assurance regarding the internal consistency of the launchpad's business
logic, it is crucial to acknowledge the inherent limitations of this
approach and outline avenues for future research. The current formal
model serves as a powerful mathematical specification, but its
relationship to the production code and the execution environment
warrants further discussion.

\subsection{6.1. The Gap Between Specification and Production
Code}\label{the-gap-between-specification-and-production-code}

A significant limitation of the current methodology is the separation
between the formally verified Dafny code and the production-level Rust
code, which is the artifact ultimately deployed to the blockchain. The
Dafny model is a pure, mathematical representation of the logic. For the
proofs to apply to the production system, there is a critical, implicit
step: a human expert must manually audit the Rust implementation to
ensure it is a faithful and precise translation of the verified Dafny
specification.

This manual verification step, while standard in many formal methods
applications, introduces a potential point of failure. Future work could
focus on bridging this ``specification-implementation gap'' to achieve
machine-checkable correspondence. Two promising directions emerge:

\begin{enumerate}
\def\labelenumi{\arabic{enumi}.}
\item
  \textbf{Verified Code Generation:} One approach is to generate the
  production code directly from the verified specification. A trusted
  ``Dafny-to-Rust'' compiler could translate the proven Dafny logic into
  a Rust module, ensuring by construction that the deployed code adheres
  to the formal model. However, this approach faces significant
  practical hurdles. At present, no production-ready and trusted
  Dafny-to-Rust generator exists. Moreover, such a tool would need to be
  highly specialized to support the specific features of the NEAR
  blockchain, including its state management and asynchronous contract
  model, which represents a substantial engineering challenge in its own
  right.
\item
  \textbf{Integrated Specification and Implementation:} An alternative
  and more modern approach involves tools that allow formal
  specifications and proofs to be written directly within the production
  code. Languages and tools like \textbf{Verus} enable annotating Rust
  code with preconditions, postconditions, and invariants, which are
  then verified in place \citep{mok2022verus}. This unifies the
  specification and implementation into a single artifact, eliminating
  the need for manual correspondence checks and bringing formal
  verification closer to the production code. Nevertheless, this
  approach also has its drawbacks. The Verus ecosystem, while promising,
  is at an earlier stage of development compared to Dafny, and its
  toolset for formal verification is currently less mature. A
  significant practical issue is the limited IDE support, which can make
  debugging complex formal specification rules a considerably more
  challenging task.
\end{enumerate}

\subsection{6.2. Abstraction from the NEAR Execution
Environment}\label{abstraction-from-the-near-execution-environment}

The current model is a purposeful abstraction away from the complexities
of the NEAR blockchain's execution environment. This was a necessary
simplification to make the verification of the complex financial logic
tractable. However, this abstraction inherently limits the scope of
properties that can be proven.

The model does not account for crucial aspects of the on-chain
environment, such as:

\begin{itemize}
\tightlist
\item
  The asynchronous, message-passing nature of cross-contract calls.
\item
  Gas mechanics and the possibility of out-of-gas failures.
\item
  Potential reentrancy vulnerabilities \citep{weir2018formal} arising
  from complex call patterns.
\end{itemize}

These factors can introduce complications that the current, purely
functional model cannot address. For instance, the withdrawal reentrancy
issue discovered during development highlighted how interactions with
the execution environment could affect security in ways not captured by
the abstract logic. A significant, albeit highly challenging, area for
future work would be to formalize the semantics of the NEAR execution
environment itself. This would enable proving properties that hold not
just in theory but also within the concrete operational context of the
blockchain.

\subsection{6.3. Expanding the Scope of Verified
Properties}\label{expanding-the-scope-of-verified-properties}

The current verification focuses primarily on critical safety
properties, such as the correctness of refund calculations and adherence
to the sale cap. In an ideal world, a fully verified contract would
guarantee a broader spectrum of properties. The following categories
represent a long-term roadmap for expanding the scope of verification:

\begin{itemize}
\tightlist
\item
  \textbf{Validity:} A global invariant stating that if the contract
  begins in a valid state, any possible sequence of transactions will
  keep it in a valid state. This is a generalization of the state
  machine integrity proofs, ensuring holistic system consistency.
\item
  \textbf{Liveness:} Guarantees that certain desirable states are always
  eventually reachable. A critical special case, often called
  \textbf{Liquidity}, is the property that a user can always, under some
  sequence of valid actions, withdraw their entitled funds from the
  contract. Proving liveness is notoriously difficult as it must account
  for potential interference from the external environment (e.g.,
  frontrunning or other adversarial actions).
\item
  \textbf{Fidelity:} This property ensures that the contract's internal
  representation of assets (e.g., token balances in its ledger) is
  always equal to the actual amount of assets cryptographically
  controlled by the contract. This would formally prove that funds
  cannot be lost or become permanently inaccessible due to bugs in the
  accounting logic.
\end{itemize}

The verification of these broader liveness and fidelity properties
constitutes a long-term research objective for the field. The preceding
discussion serves to delineate the precise scope of the guarantees
provided by the present work, positioning it as a foundational step
focused on core safety invariants, from which more comprehensive
verification efforts may proceed.

\section{7. Conclusion}\label{conclusion}

This paper has detailed the formal verification of a token sale
launchpad's core logic using Dafny. We have demonstrated that by
adopting a \textbf{compositional, bottom-up verification strategy}, it
is possible to formally reason about a system with complex, interacting
components and non-linear arithmetic \citep{chen2021compositional}.

The key achievements of this work include:

\begin{enumerate}
\def\labelenumi{\arabic{enumi}.}
\tightlist
\item
  \textbf{A Layered Proof Architecture:} Decomposing the problem from
  foundational mathematical lemmas to top-level state transitions,
  making a complex proof tractable.
\item
  \textbf{Verification of Non-Linear Arithmetic:} Proving and reusing a
  core set of lemmas for integer multiplication and division, which are
  essential for financial calculations.
\item
  \textbf{Proof of Critical Business Rules:} Formalizing and verifying
  rules such as the non-overlapping nature of discount periods.
\item
  \textbf{Mathematical Guarantee of Financial Safety:} The cornerstone
  of this work is the formal proof of \texttt{Lemma\_RefundIsSafe} and
  \texttt{Lemma\_AssetsRevert\_RoundTrip\_bounds}. Together, they
  demonstrate not only that refunds never exceed deposits, but also that
  precision loss from round-trip calculations is strictly and
  predictably bounded.
\item
  \textbf{Verified State Machine Lifecycle:} Proving the integrity of
  the contract's entire lifecycle, including user deposits, withdrawals,
  token claims with vesting, and post-sale distributions, ensuring
  predictable and correct state transitions over time.
\end{enumerate}

This work provides strong evidence that formal methods are not merely an
academic exercise but a practical and powerful tool for engineering
high-assurance financial systems, providing mathematical certainty where
traditional testing can only provide statistical confidence
\citep{woodcock2009formal, jovanovic2021foundations}.

\begin{center}\rule{0.5\linewidth}{0.5pt}\end{center}

\section{Appendix A: Formal Proofs of Foundational Integer Arithmetic
Properties}\label{appendix-a-formal-proofs-of-foundational-integer-arithmetic-properties}

The \texttt{MathLemmas} module constitutes the axiomatic foundation upon
which the entire verification hierarchy is constructed. Automated
theorem provers, including the Z3 SMT solver employed by Dafny, possess
comprehensive theories for linear arithmetic
\citep{kroening2016decision}. However, reasoning about non-linear
expressions involving multiplication and division often necessitates
explicit, programmer-provided proofs. This module furnishes these
proofs, creating a trusted library of fundamental mathematical
properties. This approach abstracts the intricacies of integer
arithmetic, thereby enabling the verification of higher-level business
logic in a more declarative and computationally tractable manner.

\begin{center}\rule{0.5\linewidth}{0.5pt}\end{center}

\subsection{\texorpdfstring{\textbf{Lemma 1: Monotonicity of Integer
Division}}{Lemma 1: Monotonicity of Integer Division}}\label{lemma-1-monotonicity-of-integer-division}

This lemma formally establishes that the integer division operation
(\(\lfloor a/b \rfloor\)) preserves the non-strict inequality relation
(\(\ge\)).

\textbf{Formal Specification (\texttt{Lemma\_Div\_Maintains\_GTE})}

\[
\forall x, y, k \in \mathbb{N} : (k > 0 \land x \ge y) \implies \lfloor x/k \rfloor \ge \lfloor y/k \rfloor
\]

\textbf{Description and Verification Strategy}

The lemma asserts that for any two natural numbers \(x\) and \(y\) where
\(x\) is greater than or equal to \(y\), dividing both by a positive
integer \(k\) will preserve this order relation. The proof implemented
in Dafny is a classic \emph{reductio ad absurdum}.

\begin{enumerate}
\def\labelenumi{\arabic{enumi}.}
\tightlist
\item
  \textbf{Hypothesis:} The proof begins by positing the negation of the
  consequent: \(\lfloor x/k \rfloor < \lfloor y/k \rfloor\). Within the
  domain of integers, this is equivalent to
  \(\lfloor x/k \rfloor + 1 \le \lfloor y/k \rfloor\).
\item
  \textbf{Derivation:} Leveraging the definition of Euclidean division,
  \(a = \lfloor a/b \rfloor \cdot b + (a \pmod b)\), the proof
  constructs a lower bound for \(y\) \citep{knuth1997art}. By
  substituting the hypothesis, we obtain:
  \(y \ge \lfloor y/k \rfloor \cdot k \ge (\lfloor x/k \rfloor + 1) \cdot k = (\lfloor x/k \rfloor \cdot k) + k\)
\item
  \textbf{Contradiction:} It is a known property that
  \(k > (x \pmod k)\). Therefore, we can deduce that
  \((\lfloor x/k \rfloor \cdot k) + k > (\lfloor x/k \rfloor \cdot k) + (x \pmod k) = x\).
  This establishes the inequality \(y > x\), which is a direct
  contradiction of the lemma's precondition \(x \ge y\).
\item
  \textbf{Conclusion:} As the initial hypothesis leads to a logical
  contradiction, it must be false. Consequently, the original
  consequent, \(\lfloor x/k \rfloor \ge \lfloor y/k \rfloor\), is proven
  to be true for all inputs satisfying the preconditions.
\end{enumerate}

\textbf{Verification Effectiveness:} By formalizing this property as a
standalone lemma, we provide the verifier with a powerful and reusable
inference rule. For any subsequent proof involving inequalities and
division, a simple invocation of this lemma suffices. This obviates the
need for the SMT solver to rediscover this non-trivial, non-linear
property within more complex logical contexts, thereby significantly
enhancing the automation, performance, and predictability of the overall
verification process.

\begin{center}\rule{0.5\linewidth}{0.5pt}\end{center}

\subsection{\texorpdfstring{\textbf{Lemma 2: Scaling by a Rational
Factor \(\geq\) 1
(\texttt{Lemma\_MulDivGreater\_From\_Scratch})}}{Lemma 2: Scaling by a Rational Factor \textbackslash geq 1 (Lemma\_MulDivGreater\_From\_Scratch)}}\label{lemma-2-scaling-by-a-rational-factor-geq-1-lemma_muldivgreater_from_scratch}

This lemma proves that scaling an integer by a rational factor \(y/k\)
(where \(y \ge k\)) results in a value no less than the original.

\textbf{Formal Specification}

\[
\forall x, y, k \in \mathbb{N} : (x > 0 \land k > 0 \land y \ge k) \implies \lfloor (x \cdot y) / k \rfloor \ge x
\]

\textbf{Description and Verification Strategy}

This lemma is instrumental in verifying that financial conversions at
stable or favorable prices do not lead to a loss of principal value. The
verification strategy is \textbf{compositional}, demonstrating the
elegance of building proofs upon previously established theorems.

\begin{enumerate}
\def\labelenumi{\arabic{enumi}.}
\tightlist
\item
  \textbf{Intermediate Premise:} The preconditions \(y \ge k\) and
  \(x > 0\) directly imply the inequality \(x \cdot y \ge x \cdot k\).
\item
  \textbf{Compositional Invocation:} The proof then applies the
  previously proven \texttt{Lemma\_Div\_Maintains\_GTE} to this
  intermediate inequality, substituting \(x \cdot y\) for its first
  parameter and \(x \cdot k\) for its second.
\item
  \textbf{Logical Deduction:} This invocation yields the statement
  \(\lfloor(x \cdot y)/k\rfloor \ge \lfloor(x \cdot k)/k\rfloor\).
\item
  \textbf{Simplification:} Given \(k > 0\), the term
  \(\lfloor(x \cdot k)/k\rfloor\) is definitionally equivalent to \(x\).
  This leads directly to the desired postcondition.
\end{enumerate}

\textbf{Verification Effectiveness:} This exemplifies an efficient,
layered verification approach. The proof reduces a complex, non-linear
problem to a straightforward application of a known monotonicity
property. This modularity not only enhances human comprehension but also
simplifies the task for the SMT solver, making the verification
near-instantaneous.

\begin{center}\rule{0.5\linewidth}{0.5pt}\end{center}

\subsection{\texorpdfstring{\textbf{Lemma 3: Strict Scaling by a
Rational Factor \(\geq\) 2
(\texttt{Lemma\_MulDivStrictlyGreater\_From\_Scratch})}}{Lemma 3: Strict Scaling by a Rational Factor \textbackslash geq 2 (Lemma\_MulDivStrictlyGreater\_From\_Scratch)}}\label{lemma-3-strict-scaling-by-a-rational-factor-geq-2-lemma_muldivstrictlygreater_from_scratch}

This lemma establishes a sufficient condition to guarantee a
\emph{strict} increase in value after scaling, providing a robust guard
against value loss due to integer division's truncating nature.

\textbf{Formal Specification}

\[
\forall x, y, k \in \mathbb{N} : (x > 0 \land k > 0 \land y \ge 2k) \implies \lfloor (x \cdot y) / k \rfloor > x
\]

\textbf{Description and Verification Strategy}

The proof recognizes that the precondition \(y > k\) is insufficient to
guarantee strict inequality. It employs the stronger condition
\(y \ge 2k\).

\begin{enumerate}
\def\labelenumi{\arabic{enumi}.}
\tightlist
\item
  \textbf{Strengthened Premise:} The proof establishes that \(y \ge 2k\)
  implies \(x \cdot y \ge x \cdot (2k) = x \cdot k + x \cdot k\). As
  \(x > 0\) and \(k > 0\), it follows that \(x \cdot k \ge k\). This
  allows the derivation of the crucial inequality
  \(x \cdot y \ge x \cdot k + k\).
\item
  \textbf{Compositional Invocation:} This inequality is the exact
  premise required by a stricter variant of the monotonicity lemma
  (\texttt{Lemma\_Div\_Maintains\_GT}), which proves
  \(a \ge b+k \implies \lfloor a/k \rfloor > \lfloor b/k \rfloor\).
  Applying this specialized lemma yields
  \(\lfloor(x \cdot y)/k\rfloor > \lfloor(x \cdot k)/k\rfloor\).
\item
  \textbf{Conclusion:} The term \(\lfloor(x \cdot k)/k\rfloor\)
  simplifies to \(x\), thus proving the postcondition.
\end{enumerate}

\textbf{Verification Effectiveness:} This lemma showcases a critical
aspect of formal methods: identifying the precise and sufficiently
strong preconditions required to guarantee a desired property. By
encapsulating this logic, we create a tool for reasoning about scenarios
where a tangible gain must be proven, such as the application of a
significant bonus.

\begin{center}\rule{0.5\linewidth}{0.5pt}\end{center}

\subsection{\texorpdfstring{\textbf{Lemmas 4 \& 5: Scaling by a Rational
Factor \(\leq\)
1}}{Lemmas 4 \& 5: Scaling by a Rational Factor \textbackslash leq 1}}\label{lemmas-4-5-scaling-by-a-rational-factor-leq-1}

These lemmas are the logical duals to the preceding two, addressing
scaling by factors less than or equal to one.

\textbf{Formal Specification}

\begin{enumerate}
\def\labelenumi{\arabic{enumi}.}
\item
  \texttt{Lemma\_MulDivLess\_From\_Scratch}:

  \(\forall x, y, k \in \mathbb{N} : (x > 0 \land y > 0 \land k \ge y) \implies \lfloor(x \cdot y) / k\rfloor \le x\)
\item
  \texttt{Lemma\_MulDivStrictlyLess\_From\_Scratch}:

  \(\forall x, y, k \in \mathbb{N} : (x > 0 \land y > 0 \land k > y) \implies \lfloor(x \cdot y) / k\rfloor < x\)
\end{enumerate}

\textbf{Description and Verification Strategy}

The proofs demonstrate both elegance and efficiency through reuse and
contradiction.

\begin{itemize}
\item
  The proof for the non-strict case (\texttt{...Less...}) is achieved by
  a clever reuse of

  \texttt{Lemma\_MulDivGreater\_From\_Scratch}. Given \(k \ge y\), it
  invokes the greater-than lemma with the roles of \(k\) and \(y\)
  interchanged.
\item
  The proof for the strict case (\texttt{...StrictlyLess...}) proceeds
  by contradiction. It assumes \(\lfloor(x \cdot y)/k\rfloor \ge x\),
  which implies \(x \cdot y \ge x \cdot k\), and for \(x > 0\), implies
  \(y \ge k\). This directly contradicts the lemma's precondition
  \(k > y\).
\end{itemize}

\textbf{Verification Effectiveness:} These proofs highlight the power of
a well-curated lemma library. Reusing existing proofs minimizes
redundant effort, while the declarative nature of the proof by
contradiction allows the SMT solver to efficiently explore the logical
space and confirm the inconsistency.

\begin{center}\rule{0.5\linewidth}{0.5pt}\end{center}

\subsection{\texorpdfstring{\textbf{Lemma 6: The Bounded Property of
Integer Division Truncation
(\texttt{Lemma\_DivMul\_Bounds})}}{Lemma 6: The Bounded Property of Integer Division Truncation (Lemma\_DivMul\_Bounds)}}\label{lemma-6-the-bounded-property-of-integer-division-truncation-lemma_divmul_bounds}

This lemma formalizes the fundamental property that integer division is
a truncating operation, which is the root cause of potential precision
loss in round-trip calculations. It proves not only that the result does
not exceed the original value but also that the ``loss'' is strictly
bounded.

\textbf{Formal Specification} \[
\forall x, y \in \mathbb{N} : (y > 0) \implies (\lfloor x/y \rfloor \cdot y \le x) \land (0 \le x - (\lfloor x/y \rfloor \cdot y) < y)
\]

\textbf{Description and Verification Strategy} The proof of this lemma
is a testament to the synergy between the programmer and the underlying
verification engine. It is established by asserting the
\textbf{Euclidean Division Theorem}, a core axiom within the SMT
solver's theory of integers:
\texttt{assert\ x\ ==\ (x\ /\ y)\ *\ y\ +\ (x\ \%\ y);}

From this axiom, both postconditions follow immediately. Since the
remainder \texttt{(x\ \%\ y)} is definitionally non-negative, \texttt{x}
must be greater than or equal to the term \texttt{(x\ /\ y)\ *\ y}. The
second property, \texttt{x\ -\ (x\ /\ y)\ *\ y\ \textless{}\ y}, follows
from the fact that \texttt{x\ \%\ y} is definitionally less than the
divisor \texttt{y}.

\textbf{Verification Effectiveness} This is a paradigmatic example of
effective formal verification. The programmer's role is not to re-prove
foundational mathematics but to strategically invoke known axioms to
guide the verifier's reasoning. By stating this single, axiomatic
assertion, we provide the solver with the necessary fact to prove the
safety and bounded loss of all round-trip financial calculations
throughout the system.

\begin{center}\rule{0.5\linewidth}{0.5pt}\end{center}

\subsection{\texorpdfstring{\textbf{Lemma 7: Lower Bound of Division
from Strict Multiplication
(\texttt{Lemma\_DivLowerBound\_from\_StrictMul})}}{Lemma 7: Lower Bound of Division from Strict Multiplication (Lemma\_DivLowerBound\_from\_StrictMul)}}\label{lemma-7-lower-bound-of-division-from-strict-multiplication-lemma_divlowerbound_from_strictmul}

This lemma proves a subtle but powerful property of non-linear integer
arithmetic that is often non-trivial for SMT solvers to deduce on their
own. It establishes a lower bound for a division's result based on a
strict inequality involving a product.

\textbf{Formal Specification}

\[
\forall a, b, c \in \mathbb{N} : (c > 0 \land a > b \cdot c) \implies \lfloor a/c \rfloor \ge b
\]

\textbf{Description and Verification Strategy}

The lemma asserts that if a number \texttt{a} is strictly greater than a
product \texttt{b\ *\ c}, then dividing \texttt{a} by \texttt{c} must
yield a result of at least \texttt{b}. The proof is a classic
\emph{reductio ad absurdum}.

\begin{enumerate}
\def\labelenumi{\arabic{enumi}.}
\tightlist
\item
  \textbf{Hypothesis:} The proof begins by assuming the negation of the
  consequent: \(\lfloor a/c \rfloor < b\). For integers, this is
  equivalent to \(\lfloor a/c \rfloor \le b - 1\).
\item
  \textbf{Derivation:} Using the definition of Euclidean division,
  \(a = \lfloor a/c \rfloor \cdot c + (a \pmod c)\), the proof
  constructs an upper bound for \texttt{a}. By substituting the
  hypothesis, we obtain:
  \(a \le (b - 1) \cdot c + (a \pmod c) = b \cdot c - c + (a \pmod c)\)
\item
  \textbf{Contradiction:} We know that the remainder \((a \pmod c)\) is
  strictly less than the divisor \texttt{c}. This allows us to establish
  a strict inequality:
  \(b \cdot c - c + (a \pmod c) < b \cdot c - c + c = b \cdot c\). This
  establishes that \(a < b \cdot c\), which is a direct contradiction of
  the lemma's precondition \(a > b \cdot c\).
\item
  \textbf{Conclusion:} As the initial hypothesis leads to a logical
  contradiction, it must be false. Consequently, the original
  consequent, \(\lfloor a/c \rfloor \ge b\), is proven to be true.
\end{enumerate}

\textbf{Verification Effectiveness}

This lemma serves as a crucial piece of guidance for the verifier. By
proving this non-linear property explicitly, we equip the SMT solver
with a ready-made inference rule. This is particularly vital in proofs
of round-trip calculations with tight bounds, such as
\texttt{Lemma\_WeightOriginal\_RoundTrip\_bounds}, where proving that a
value is greater than or equal to \texttt{amount\ -\ 1} requires exactly
this kind of reasoning. Encapsulating this logic prevents the solver
from getting stuck or timing out while trying to rediscover this
relationship in a more complex context, thereby improving the robustness
and performance of the overall verification.

\section{Appendix B: Formal Verification of Asset Conversion
Logic}\label{appendix-b-formal-verification-of-asset-conversion-logic}

The \texttt{AssetCalculations} module represents the first layer of
application-specific business logic, constructed upon the axiomatic
foundation established in \texttt{MathLemmas}. Its purpose is to
translate the abstract mathematical properties of integer arithmetic
into concrete, provable guarantees for financial asset conversion
operations. This module defines the pure mathematical specifications for
conversion and provides a comprehensive suite of lemmas that formally
prove their key properties, such as monotonicity, predictable behavior
under various price conditions, and, most critically, round-trip safety.

\subsection{B.1. Core Specification
Functions}\label{b.1.-core-specification-functions}

At the heart of the module lie two pure functions defining the
mathematical essence of forward and reverse conversion. For clarity, let
\(w \in \mathbb{N}\) represent the input amount (weight or principal),
\(d_T \in \mathbb{N}^+\) be the denominator of the price fraction (e.g.,
the deposit token amount), and \(s_T \in \mathbb{N}^+\) be the numerator
(e.g., the sale token amount). Let \(C\) denote the direct conversion
(\texttt{CalculateAssetsSpec}) and \(R\) denote the reverse conversion
(\texttt{CalculateAssetsRevertSpec}).

\begin{enumerate}
\def\labelenumi{\arabic{enumi}.}
\item
  \textbf{Direct Conversion (\(C\)):} This function maps a principal
  amount into a quantity of target assets.
  \[ C(w, d_T, s_T) := \lfloor (w \cdot s_T) / d_T \rfloor \]
\item
  \textbf{Reverse Conversion (\(R\)):} This function performs the
  inverse operation, calculating the principal amount from a quantity of
  assets. \[ R(w, d_T, s_T) := \lfloor (w \cdot d_T) / s_T \rfloor \]
\end{enumerate}

\subsection{\texorpdfstring{B.2. Verification of Direct Conversion
Properties
(\texttt{CalculateAssets})}{B.2. Verification of Direct Conversion Properties (CalculateAssets)}}\label{b.2.-verification-of-direct-conversion-properties-calculateassets}

This group of lemmas proves intuitive economic properties of the
function \(C\) by directly mapping them to the foundational lemmas from
Appendix A.

\textbf{Lemma B.2.1: Conversion at a Non-Disadvantageous Price
(\texttt{Lemma\_CalculateAssets\_IsGreaterOrEqual})}

\begin{itemize}
\tightlist
\item
  \textbf{Formal Specification:}
  \[ \forall w, d_T, s_T \in \mathbb{N}^+ : (s_T \ge d_T) \implies C(w, d_T, s_T) \ge w \]
\item
  \textbf{Description and Verification Strategy:} This lemma guarantees
  that if the exchange rate is stable or favorable ( \(s_T \ge d_T\)),
  the resulting asset quantity has a nominal value no less than the
  original principal. The proof is a direct \textbf{instantiation} of
  \texttt{Lemma\_MulDivGreater\_From\_Scratch}. The parameters are
  mapped as follows: \(x \to w\), \(y \to s_T\), \(k \to d_T\). The
  lemma's precondition \(s_T \ge d_T\) precisely matches the required
  precondition \(y \ge k\) from \texttt{MathLemmas}.
\item
  \textbf{Verification Effectiveness:} This demonstrates the power of
  compositional reasoning. A complex financial guarantee is proven with
  a single invocation of a previously verified, general-purpose lemma,
  making the proof trivial for the SMT solver and transparent to a human
  auditor.
\end{itemize}

\textbf{Lemma B.2.2: Conversion at a Highly Advantageous Price
(\texttt{Lemma\_CalculateAssets\_IsGreater})}

\begin{itemize}
\tightlist
\item
  \textbf{Formal Specification:}
  \[ \forall w, d_T, s_T \in \mathbb{N}^+ : (s_T \ge 2 \cdot d_T) \implies C(w, d_T, s_T) > w \]
\item
  \textbf{Description and Verification Strategy:} This guarantees a
  \emph{strict} increase in nominal value when the exchange rate is
  significantly favorable. The precondition \(s_T \ge 2 \cdot d_T\) is
  necessary to overcome the truncating effect of integer division. The
  proof is a direct instantiation of
  \texttt{Lemma\_MulDivStrictlyGreater\_From\_Scratch}.
\item
  \textbf{Verification Effectiveness:} This highlights the importance of
  identifying precise preconditions to obtain strict guarantees. The
  lemma is crucial for proving scenarios where not just non-loss, but a
  tangible gain, must be formally assured.
\end{itemize}

\textbf{Lemma B.2.3: Conversion at an Unfavorable Price
(\texttt{Lemma\_CalculateAssets\_IsLess})}

\begin{itemize}
\item
  \textbf{Formal Specification:}

  \(\forall w, d_T, s_T \in \mathbb{N}^+ : (s_T < d_T) \implies C(w, d_T, s_T) < w\)
\item
  \textbf{Description and Verification Strategy:} This guarantees that
  if the exchange rate is unfavorable, the resulting asset quantity has
  a nominal value strictly less than the original principal. The proof
  is a direct instantiation of
  \texttt{Lemma\_MulDivStrictlyLess\_From\_Scratch}.
\item
  \textbf{Verification Effectiveness:} This completes the suite of
  behavioral guarantees for the \(C\) function, covering all three
  possible price relationships (\(\ge\), \(=\), \(<\)) and ensuring the
  function's behavior is fully specified and proven.
\end{itemize}

\subsection{\texorpdfstring{B.3. Verification of Reverse Conversion
Properties
(\texttt{CalculateAssetsRevert})}{B.3. Verification of Reverse Conversion Properties (CalculateAssetsRevert)}}\label{b.3.-verification-of-reverse-conversion-properties-calculateassetsrevert}

This set of lemmas proves symmetric properties for the reverse function
\(R\). The verification strategy is analogous: instantiation of
foundational lemmas. The key observation is that
\texttt{R(w,\ d\_T,\ s\_T)} is mathematically equivalent to
\texttt{C(w,\ s\_T,\ d\_T)}, meaning a reverse conversion is simply a
direct conversion with the roles of the price fraction's numerator and
denominator exchanged.

\textbf{Lemma B.3.1: Reversion from an Originally Unfavorable Price}

\textbf{(\texttt{Lemma\_CalculateAssetsRevert\_IsGreaterOrEqual})}

\begin{itemize}
\tightlist
\item
  \textbf{Formal Specification:}
  \[ \forall w, d_T, s_T \in \mathbb{N}^+ : (d_T \ge s_T) \implies R(w, d_T, s_T) \ge w \]
\item
  \textbf{Description and Verification Strategy:} If the original price
  was unfavorable or stable for the user (\(s_T \le d_T\)), then
  converting the assets back will yield a principal amount no less than
  the asset amount being converted. The proof invokes
  \texttt{Lemma\_MulDivGreater\_From\_Scratch} with the parameter
  mapping \(x \to w\), \(y \to d_T\), \(k \to s_T\). The precondition
  \(d_T \ge s_T\) correctly satisfies the required \(y \ge k\).
\item
  \textbf{Verification Effectiveness:} This demonstrates the elegance of
  symmetric arguments in formal proofs. Instead of constructing a new
  complex proof, we reuse an existing lemma by simply permuting its
  arguments, which serves to validate the generality and correctness of
  the foundational axioms.
\end{itemize}

\subsection{B.4. Verification of Composite and Crucial Safety
Properties}\label{b.4.-verification-of-composite-and-crucial-safety-properties}

These lemmas establish higher-order properties that are critical for the
overall safety and integrity of the financial logic.

\textbf{Lemma B.4.1: Monotonicity of Reverse Conversion
(\texttt{Lemma\_CalculateAssetsRevertSpec\_Monotonic})}

\begin{itemize}
\item
  \textbf{Formal Specification:}

  \(\forall w_1, w_2, d_T, s_T \in \mathbb{N}^+ : (w_1 \le w_2) \implies R(w_1, d_T, s_T) \le R(w_2, d_T, s_T)\)
\item
  \textbf{Description and Verification Strategy:} This lemma proves that
  the reverse conversion function \(R\) is monotonic. That is,
  converting a smaller quantity of assets back to the principal cannot
  yield a larger result than converting a larger quantity. This property
  is an absolute prerequisite for proving the safety of partial refund
  calculations. The proof is based on
  \texttt{Lemma\_Div\_Maintains\_GTE}. From \(w_1 \le w_2\), it follows
  that \(w_1 \cdot d_T \le w_2 \cdot d_T\). Applying
  \texttt{Lemma\_Div\_Maintains\_GTE} to this inequality with divisor
  \(s_T\) directly yields the desired consequent.
\item
  \textbf{Verification Effectiveness:} This shows how foundational
  lemmas are used to prove higher-order properties ( monotonicity),
  which in turn serve as essential building blocks for even more complex
  safety proofs, such as refund correctness.
\end{itemize}

\textbf{B.4.2: The Algebraic Equation for Round-Trip Loss
(\texttt{Lemma\_RoundTripLossEquation})}

\begin{itemize}
\tightlist
\item
  \textbf{Formal Specification:} Let \(w, d_T, s_T \in \mathbb{N}^+\).
  Let:

  \begin{itemize}
  \tightlist
  \item
    \(assets := \lfloor (w \cdot s_T) / d_T \rfloor\)
  \item
    \(reverted := \lfloor (assets \cdot d_T) / s_T \rfloor\)
  \item
    \(rem_1 := (w \cdot s_T) \pmod{d_T}\)
  \item
    \(rem_2 := (assets \cdot d_T) \pmod{s_T}\) Then the following
    equality holds: \[ (w - reverted) \cdot s_T = rem_1 + rem_2 \]
  \end{itemize}
\item
  \textbf{Description and Verification Strategy:} This lemma provides
  the algebraic foundation for proving the bounded loss property. It
  isolates the complex arithmetic into a single, elegant equation. It
  proves that the ``loss'' from a round-trip conversion, when scaled up
  by \texttt{sT}, is precisely equal to the sum of the remainders from
  the two integer division operations involved. The proof in Dafny is a
  straightforward algebraic manipulation using the \texttt{calc}
  statement, which makes the reasoning explicit and easy for the
  verifier to follow:

  \begin{enumerate}
  \def\labelenumi{\arabic{enumi}.}
  \tightlist
  \item
    Start with the expression \texttt{(w\ -\ reverted)\ *\ sT}.
  \item
    Apply the Euclidean division theorem to \texttt{w\ *\ sT},
    substituting it with \texttt{(assets\ *\ dT\ +\ rem1)}.
  \item
    Similarly, apply the theorem to \texttt{assets\ *\ dT}, substituting
    it with \texttt{(reverted\ *\ sT\ +\ rem2)}.
  \item
    The expression becomes
    \texttt{((reverted\ *\ sT\ +\ rem2)\ +\ rem1)\ -\ (reverted\ *\ sT)}.
  \item
    Simplifying this expression directly yields \texttt{rem1\ +\ rem2},
    completing the proof.
  \end{enumerate}
\item
  \textbf{Verification Effectiveness:} This lemma is a prime example of
  effective proof engineering. By isolating this non-trivial algebraic
  identity into a standalone proof, we greatly simplify the main safety
  proof of \texttt{Lemma\_AssetsRevert\_RoundTrip\_bounds}. Instead of
  forcing the SMT solver to re-derive this equality from first
  principles within a more complex logical context, we provide it as a
  trusted, reusable theorem. This makes the final proof more readable,
  robust, and computationally efficient.
\end{itemize}

\textbf{Lemma B.4.3: Round-Trip Calculation Safety and Bounded Loss
(\texttt{Lemma\_AssetsRevert\_RoundTrip\_bounds})}

\begin{itemize}
\item
  \textbf{Formal Specification:}

  \(\forall w, d_T, s_T \in \mathbb{N}^+ : C(w, d_T, s_T) > 0 \implies R(C(w, d_T, s_T), d_T, s_T) \le w \land (w - R(C(w, d_T, s_T), d_T, s_T)) \cdot s_T < d_T + s_T\)
\item
  \textbf{Description and Verification Strategy:} This is the
  \textbf{central safety guarantee} of this module. It formally proves
  that the sequential application of a direct conversion and a reverse
  conversion cannot create value \emph{ex nihilo}. Furthermore, it
  proves the stronger property that the value loss from integer
  truncation is strictly bounded. The proof is a composition of several
  established facts:

  \begin{enumerate}
  \def\labelenumi{\arabic{enumi}.}
  \tightlist
  \item
    Let \texttt{assets\ :=\ C(w,\ d\_T,\ s\_T)} and
    \texttt{reverted\ :=\ R(assets,\ d\_T,\ s\_T)}.
  \item
    The \texttt{reverted\ \textless{}=\ w} inequality is proven as
    before, using \texttt{Lemma\_DivMul\_Bounds} and
    \texttt{Lemma\_Div\_Maintains\_GTE}.
  \item
    The stronger bounded loss property is proven using a dedicated
    helper lemma, \texttt{Lemma\_RoundTripLossEquation}. This lemma
    algebraically demonstrates that the scaled loss,
    \texttt{(w\ -\ reverted)\ *\ sT}, is exactly equal to the sum of the
    remainders from the two division operations:
    \texttt{(w\ *\ sT)\ \%\ dT\ +\ (assets\ *\ dT)\ \%\ sT}.
  \item
    Since a remainder from division by \texttt{k} is always strictly
    less than \texttt{k}, we know that
    \texttt{(w\ *\ sT)\ \%\ dT\ \textless{}\ dT} and
    \texttt{(assets\ *\ dT)\ \%\ sT\ \textless{}\ sT}.
  \item
    Summing these two inequalities gives
    \texttt{rem1\ +\ rem2\ \textless{}\ dT\ +\ sT}, which completes the
    proof.
  \end{enumerate}
\item
  \textbf{Verification Effectiveness:} This lemma is the culmination of
  the \texttt{AssetCalculations} module. It demonstrates how multiple
  simple, proven properties can be chained to prove a complex,
  critically important safety property. It provides not just a guarantee
  against value creation, but a strict, provable upper bound for any
  truncation-related losses.
\end{itemize}

\section{Appendix C: Formal Verification of Time-Based Discount
Logic}\label{appendix-c-formal-verification-of-time-based-discount-logic}

The \texttt{Discounts} module formalizes the logic for applying
time-sensitive percentage-based bonuses. It employs fixed-point
arithmetic to handle percentages with precision and establishes a
rigorous framework to ensure that discount rules are applied
consistently and unambiguously. The verification effort for this module
guarantees not only the correctness of the core financial calculations
but also the logical integrity of collections of discounts, preventing
common business logic flaws such as applying multiple bonuses
simultaneously.

\subsection{C.1. Foundational Definitions and
Predicates}\label{c.1.-foundational-definitions-and-predicates}

The module is built upon a set of core definitions representing the
properties of a single discount. Let the constant \(M\) denote the
\texttt{MULTIPLIER} (e.g., 10000 for four decimal places of precision),
which serves as the basis for fixed-point arithmetic. A
\texttt{Discount}, \(d\), is a tuple \((s, e, p)\) where
\(s, e, p \in \mathbb{N}\), representing \texttt{startDate},
\texttt{endDate}, and \texttt{percentage} respectively.

\textbf{Predicate C.1.1: Validity of a Discount
(\texttt{ValidDiscount})}

\begin{itemize}
\tightlist
\item
  \textbf{Formal Specification:} A discount \(d = (s, e, p)\) is
  considered valid if its parameters are self-consistent.
  \[ \text{ValidDiscount}(d) \iff (p \in (0, M] \land s < e) \]
\item
  \textbf{Description:} This predicate enforces two fundamental business
  rules: the discount percentage \(p\) must be positive and not exceed
  100\% (represented by \(M\)), and the time interval must be logical
  (the start date must precede the end date). This predicate forms the
  base assumption for all operations on a discount.
\end{itemize}

\textbf{Predicate C.1.2: Activity of a Discount (\texttt{IsActive})}

\begin{itemize}
\tightlist
\item
  \textbf{Formal Specification:} A discount \(d = (s, e, p)\) is active
  at a given time \(t \in \mathbb{N}\) if \(t\) falls within its
  effective time range. \[ \text{IsActive}(d, t) \iff s \le t < e \]
\item
  \textbf{Description:} This defines the discount's active period as a
  half-open interval \([s, e)\). This is a common and unambiguous
  convention in time-based systems, ensuring that \texttt{endDate} is
  the first moment in time when the discount is no longer active.
\end{itemize}

\subsection{C.2. Verification of Discount Application
Logic}\label{c.2.-verification-of-discount-application-logic}

This section formalizes the application of a discount to a principal
amount and proves its mathematical properties. Let \(W_A(a, p)\) denote
the \texttt{CalculateWeightedAmount} function, where
\(a \in \mathbb{N}^+\) is the amount and \(p\) is the percentage from a
valid discount.

\textbf{Function C.2.1: Weighted Amount Calculation
(\texttt{CalculateWeightedAmount})}

\begin{itemize}
\tightlist
\item
  \textbf{Formal Specification:}
  \[ W_A(a, p) := \lfloor (a \cdot (M + p)) / M \rfloor \]
\item
  \textbf{Description:} This function calculates the new ``weighted''
  amount by scaling the original amount \(a\) by a factor of
  \((1 + p/M)\). The formula is implemented using integer arithmetic to
  avoid floating-point numbers.
\end{itemize}

\textbf{Lemma C.2.2: Non-Decreasing Property of Discount Application}

\textbf{(\texttt{Lemma\_CalculateWeightedAmount\_IsGreaterOrEqual})}

\begin{itemize}
\tightlist
\item
  \textbf{Formal Specification:}
  \[ \forall a, p \in \mathbb{N}^+ : W_A(a, p) \ge a \]
\item
  \textbf{Description and Verification Strategy:} This lemma guarantees
  that applying any valid discount will never decrease the principal
  amount. The proof is a direct instantiation of
  \texttt{Lemma\_MulDivGreater\_From\_Scratch} from Appendix A. Since
  \(p > 0\), it holds that \(M + p \ge M\). This satisfies the
  \(y \ge k\) precondition, making the proof trivial.
\end{itemize}

\subsection{C.3. Verification of Discount Reversion
Logic}\label{c.3.-verification-of-discount-reversion-logic}

This section handles the inverse operation: calculating the original
amount from a weighted amount. Let \(O_A(w_a, p)\) denote
\texttt{CalculateOriginalAmount}, where \(w_a \in \mathbb{N}^+\) is the
weighted amount.

\textbf{Function C.3.1: Original Amount Calculation
(\texttt{CalculateOriginalAmount})}

\begin{itemize}
\tightlist
\item
  \textbf{Formal Specification:}
  \[ O_A(w_a, p) := \lfloor (w_a \cdot M) / (M + p) \rfloor \]
\item
  \textbf{Description:} This function reverts the discount application,
  effectively scaling the weighted amount \(w_a\) by a factor of
  \(M / (M + p)\).
\end{itemize}

\textbf{Lemma C.3.2: Non-Increasing Property of Discount Reversion}

\textbf{(\texttt{Lemma\_CalculateOriginalAmount\_IsLessOrEqual})}

\begin{itemize}
\tightlist
\item
  \textbf{Formal Specification:}
  \[ \forall w_a, p \in \mathbb{N}^+ : O_A(w_a, p) \le w_a \]
\item
  \textbf{Description and Verification Strategy:} This guarantees that
  reverting a discount cannot result in a value greater than the
  weighted amount it was derived from. The proof instantiates
  \texttt{Lemma\_MulDivLess\_From\_Scratch}. Since \(p > 0\), it holds
  that \(M \le M + p\), which satisfies the \(k \ge y\) precondition.
\end{itemize}

\subsection{C.4. Verification of Collection Consistency
Properties}\label{c.4.-verification-of-collection-consistency-properties}

These properties are critical as they govern the behavior of a set of
discounts, ensuring logical integrity at the system level. Let
\(D = (d_0, d_1, ..., d_{n-1})\) be a sequence of discounts.

\textbf{Predicate C.4.1: Non-Overlapping Discounts
(\texttt{DiscountsDoNotOverlap})}

\begin{itemize}
\item
  \textbf{Formal Specification:} A sequence of discounts \(D\) is
  non-overlapping if for any two distinct discounts, their active time
  intervals are disjoint. Let \(d_i = (s_i, e_i, p_i)\).

  \(\text{DiscountsDoNotOverlap}(D) \iff \forall i, j \in [0, n-1] : (i < j \implies e_i \le s_j \lor e_j \le s_i)\)
\item
  \textbf{Description:} This is a crucial business rule that prevents
  ambiguity. It ensures that no two discount periods can be active at
  the same time, which is fundamental for deterministic calculations.
\end{itemize}

\textbf{Lemma C.4.2: Uniqueness of Active Discount
(\texttt{Lemma\_UniqueActiveDiscount})}

\begin{itemize}
\item
  \textbf{Formal Specification:} If a sequence of discounts \(D\) is
  non-overlapping, then at any given time \(t\), at most one discount in
  the sequence can be active.

  \(\text{DiscountsDoNotOverlap}(D) \implies \forall i, j \in [0, n-1], \forall t \in \mathbb{N} : (\text{IsActive}(d_i, t) \land \text{IsActive}(d_j, t) \implies i = j)\)
\item
  \textbf{Description and Verification Strategy:} This is the most
  important safety property for the collection of discounts. It
  guarantees that any search for an active discount will yield an
  unambiguous result. The proof proceeds by contradiction. Assume
  \(i \ne j\) and both \(d_i\) and \(d_j\) are active at time \(t\).

  \begin{enumerate}
  \def\labelenumi{\arabic{enumi}.}
  \tightlist
  \item
    \(\text{IsActive}(d_i, t) \implies s_i \le t < e_i\)
  \item
    \(\text{IsActive}(d_j, t) \implies s_j \le t < e_j\)
  \item
    From these, it follows that \(s_i < e_j\) and \(s_j < e_i\).
  \item
    This contradicts the \texttt{DiscountsDoNotOverlap(D)} predicate,
    which requires \(e_i \le s_j\) or \(e_j \le s_i\).
  \item
    Therefore, the initial assumption (\(i \ne j\)) must be false,
    proving that \(i = j\).
  \end{enumerate}
\item
  \textbf{Verification Effectiveness:} This lemma is a prime example of
  proving a high-level system property as a direct logical consequence
  of a lower-level data invariant. By verifying this, Dafny provides a
  mathematical guarantee that the core business logic for finding and
  applying bonuses is free from race conditions or ambiguity related to
  time, which is a common and critical failure mode in financial systems
  \citep{luu2016making}.
\end{itemize}

\section{Appendix D: Formal Verification of System Configuration and
Composite
Logic}\label{appendix-d-formal-verification-of-system-configuration-and-composite-logic}

The \texttt{Config} module serves as the central nervous system of the
launchpad specification. It aggregates all system parameters, business
rules, and component configurations into a single, immutable data
structure. This module's primary function is to compose the verified
primitives from lower-level modules (such as \texttt{Discounts}) into
higher-level, context-aware specifications. Its verification ensures
that these composite operations maintain the safety properties
established by their constituent parts and that the system's overall
parameterization is logically sound \citep{almeida2007compositional}.

\subsection{\texorpdfstring{D.1. The \texttt{Config} Datatype and Core
Invariants}{D.1. The Config Datatype and Core Invariants}}\label{d.1.-the-config-datatype-and-core-invariants}

The state of the system's static configuration is captured by the
\texttt{Config} datatype, denoted here as \(\Gamma\). It is a tuple
comprising various parameters, including the sale mechanics, dates, and
sequences of sub-structures like discounts.

\textbf{Predicate D.1.1: System-Wide Validity (\texttt{ValidConfig})}

The \texttt{ValidConfig} predicate is the root invariant for the entire
system. It asserts that the configuration \(\Gamma\) is well-formed by
taking a logical conjunction of numerous component-level validity
predicates, ensuring holistic integrity before any transaction is
processed.

\begin{itemize}
\item
  \textbf{Formal Specification:} Let \(\Gamma\) be a configuration
  instance.

  \(\text{ValidConfig}(\Gamma) \iff P_{\text{dates}} \land P_{\text{mechanics}} \land P_{\text{discounts}} \land P_{\text{vesting}} \land P_{\text{stakeholders}} \land P_{\text{accounting}}\)
  where each component predicate is defined as:

  \begin{itemize}
  \item
    \textbf{Date Consistency (\(P_{\text{dates}}\)):}
    \(\Gamma.startDate < \Gamma.endDate\)
  \item
    \textbf{Mechanics Consistency (\(P_{\text{mechanics}}\)):}

    \(\Gamma.mechanic.FixedPrice? \implies (\Gamma.mechanic.depositTokenAmount > 0 \land \Gamma.mechanic.saleTokenAmount > 0)\)
  \item
    \textbf{Discounts Consistency (\(P_{\text{discounts}}\)):}
    \(DiscountsDoNotOverlap(\Gamma.discount) \land (\forall d \in \Gamma.discount : ValidDiscount(d))\)
  \item
    \textbf{Global Vesting Consistency (\(P_{\text{vesting}}\)):}

    \(\Gamma.vestingSchedule.Some? \implies ValidVestingSchedule(\Gamma.vestingSchedule.v)\)
  \item
    \textbf{Stakeholder Consistency (\(P_{\text{stakeholders}}\)):}

    \(\text{IsUnique}(\Gamma.distributionProportions) \land (\forall p \in \Gamma.distributionProportions.stakeholderProportions : p.Valid())\)
  \item
    \textbf{Accounting Consistency (\(P_{\text{accounting}}\)):}

    \(\Gamma.totalSaleAmount = \Gamma.saleAmount + \text{SumOfStakeholderAllocations}(\Gamma.distributionProportions)\)
  \end{itemize}
\item
  \textbf{Description:} This predicate establishes a comprehensive
  baseline of sanity for the system's parameters. In addition to basic
  checks on dates and sale mechanics, it now enforces several critical
  invariants:

  \begin{itemize}
  \tightlist
  \item
    \textbf{Accounting Integrity:} It guarantees that the
    \texttt{totalSaleAmount} is precisely the sum of the public sale
    amount and all private stakeholder allocations. This prevents
    configuration-level bugs that could lead to token supply inflation
    or deflation.
  \item
    \textbf{Stakeholder Uniqueness:} It ensures that all stakeholder
    accounts (including the solver) are unique, preventing ambiguous or
    incorrect distributions.
  \item
    \textbf{Recursive Validity:} It recursively validates each
    individual stakeholder's configuration, including any private
    vesting schedules they may have.
  \end{itemize}

  \texttt{ValidConfig} serves as a crucial precondition for all
  functions that operate on the configuration, ensuring they are never
  invoked with inconsistent or illogical data.
\end{itemize}

\subsection{D.2. High-Level Specification of Composite
Calculations}\label{d.2.-high-level-specification-of-composite-calculations}

This section analyzes the core functions within \texttt{Config} that
combine the system's state (time) with financial primitives (discount
application) to produce context-dependent results.

\textbf{Function D.2.1: Specification for Weighted Amount Calculation
(\texttt{CalculateWeightedAmountSpec})}

Let \(W_S(a, t, \Gamma)\) denote this specification, which computes the
weighted amount for a principal \(a\) at time \(t\) under configuration
\(\Gamma\). Let \(F(D, t)\) be the \texttt{FindActiveDiscountSpec}
function, which returns \texttt{Some(d)} if an active discount \(d\)
exists in sequence \(D\) at time \(t\), and \texttt{None} otherwise.

\begin{itemize}
\tightlist
\item
  \textbf{Formal Specification:} For \(a > 0\): \[
  W_S(a, t, \Gamma) := \begin{cases}
  a & \text{if } F(\Gamma.\text{discount}, t) = \text{None} \\
  W_A(a, d.p) & \text{if } F(\Gamma.\text{discount}, t) = \text{Some}(d)
  \end{cases}
  \] where \(W_A(a, p)\) is the \texttt{CalculateWeightedAmount}
  function from Appendix C.
\item
  \textbf{Description:} This function acts as a logical switch. It
  models the behavior of applying a discount if and only if one is
  active at the specified time. It encapsulates the search-and-apply
  logic into a single pure function.
\end{itemize}

\textbf{Lemma D.2.2: Monotonicity of Weighted Amount Calculation}

\textbf{(\texttt{Lemma\_CalculateWeightedAmountSpec\_Monotonic})}

\begin{itemize}
\tightlist
\item
  \textbf{Formal Specification:} \[
  \forall a_1, a_2, t \in \mathbb{N}, \forall \Gamma : (\text{ValidConfig}(\Gamma) \land a_1 \le a_2) \implies W_S(a_1, t, \Gamma) \le W_S(a_2, t, \Gamma)
  \]
\item
  \textbf{Description and Verification Strategy:} This critical lemma
  proves that the system-level weighting function is monotonic. The
  proof proceeds by case analysis on the result of
  \(F(\Gamma.discount, t)\):

  \begin{enumerate}
  \def\labelenumi{\arabic{enumi}.}
  \tightlist
  \item
    \textbf{Case \texttt{None}:} \(W_S(a, t, \Gamma) = a\). The property
    reduces to \(a_1 \le a_2\), which is true by the precondition.
  \item
    \textbf{Case \texttt{Some(d)}:} The property becomes
    \(W_A(a_1, d.p) \le W_A(a_2, d.p)\). This is equivalent to proving
    \(\lfloor(a_1 \cdot (M+p))/M\rfloor \le \lfloor(a_2 \cdot (M+p))/M\rfloor\).
    Given \(a_1 \le a_2\), it follows that
    \(a_1 \cdot (M+p) \le a_2 \cdot (M+p)\). Applying
    \texttt{Lemma\_Div\_Maintains\_GTE} from Appendix A completes the
    proof for this case.
  \end{enumerate}
\item
  \textbf{Verification Effectiveness:} This lemma is essential for
  reasoning about aggregate values in the system, such as total
  deposits. It provides a formal guarantee that larger initial
  contributions will always result in equal or larger weighted
  contributions, a fundamental property for fairness.
\end{itemize}

\subsection{D.3. Ultimate Round-Trip Safety for Composite
Logic}\label{d.3.-ultimate-round-trip-safety-for-composite-logic}

This section culminates in proving the round-trip safety for the entire
chain of time-dependent bonus calculations.

\textbf{Function D.3.1: Specification for Original Amount Calculation
(\texttt{CalculateOriginalAmountSpec})}

Let \(O_S(w_a, t, \Gamma)\) denote this specification for a weighted
amount \(w_a\).

\begin{itemize}
\tightlist
\item
  \textbf{Formal Specification:} For \(w_a > 0\): \[
  O_S(w_a, t, \Gamma) := \begin{cases}
  w_a & \text{if } F(\Gamma.\text{discount}, t) = \text{None} \\
  O_A(w_a, d.p) & \text{if } F(\Gamma.\text{discount}, t) = \text{Some}(d)
  \end{cases}
  \] where \(O_A\) is the \texttt{CalculateOriginalAmount} function from
  Appendix C.
\end{itemize}

\textbf{Lemma D.3.2: Monotonicity of Original Amount Calculation}

\textbf{(\texttt{Lemma\_CalculateOriginalAmountSpec\_Monotonic})}

\begin{itemize}
\tightlist
\item
  \textbf{Formal Specification:} \[
  \forall r_1, r_2, t \in \mathbb{N}, \forall \Gamma : (\text{ValidConfig}(\Gamma) \land r_1 \le r_2) \implies O_S(r_1, t, \Gamma) \le O_S(r_2, t, \Gamma)
  \]
\item
  \textbf{Description and Verification Strategy:} This lemma proves that
  the system-level discount reversion function is monotonic. It is the
  logical dual to \texttt{Lemma\_CalculateWeightedAmountSpec\_Monotonic}
  and guarantees that reverting a smaller weighted amount cannot yield a
  larger original amount than reverting a larger weighted amount. This
  property is an indispensable link in the chain of reasoning for
  proving refund safety. The proof proceeds by a case analysis on the
  presence of an active discount:

  \begin{enumerate}
  \def\labelenumi{\arabic{enumi}.}
  \tightlist
  \item
    \textbf{Case \texttt{None}:} The function is an identity, so the
    property reduces to the precondition \(r_1 \le r_2\).
  \item
    \textbf{Case \texttt{Some(d)}:} The property reduces to proving
    \(\lfloor(r_1 \cdot M)/(M+p)\rfloor \le \lfloor(r_2 \cdot M)/(M+p)\rfloor\).
    Given \(r_1 \le r_2\), it follows that
    \(r_1 \cdot M \le r_2 \cdot M\). Applying the foundational
    \texttt{Lemma\_Div\_Maintains\_GTE} from Appendix A directly
    completes the proof.
  \end{enumerate}
\item
  \textbf{Verification Effectiveness:} This lemma is essential for
  composing the high-level safety proof of \texttt{Lemma\_RefundIsSafe}.
  It allows the verifier to transitively reason about inequalities.
  Specifically, it enables the proof to carry the bound established on
  the intermediate \texttt{remain} variable (which is a weighted amount)
  back to the final \texttt{refund} variable (which is an original
  amount), thus completing the deductive chain.
\end{itemize}

\begin{center}\rule{0.5\linewidth}{0.5pt}\end{center}

\textbf{Lemma D.3.3: System-Level Round-Trip Safety with Bounded Loss of
At Most One (\texttt{Lemma\_WeightOriginal\_RoundTrip\_bounds}) }

This is a paramount safety property proven within the \texttt{Config}
module. It ensures that the composite operation of applying and then
reverting a time-based discount is non-value-creating and, more
importantly, has an extremely small and strictly bounded precision loss.

\begin{itemize}
\item
  \textbf{Formal Specification:}

  \(\forall a \in \mathbb{N}^+, \forall t \in \mathbb{N}, \forall \Gamma : \text{ValidConfig}(\Gamma) \implies a - 1 \le O_S(W_S(a, t, \Gamma), t, \Gamma) \le a\)
\item
  \textbf{Description and Verification Strategy:} This proof provides
  one of the strongest guarantees in the system. It confirms that after
  applying a bonus and then reverting it, the final amount can be at
  most \textbf{one single minimal unit of currency} less than the
  original. This is achieved by a precise analysis of integer division
  truncation. The proof again proceeds by case analysis on
  \(F(\Gamma.discount, t)\):

  \begin{enumerate}
  \def\labelenumi{\arabic{enumi}.}
  \tightlist
  \item
    \textbf{Case \texttt{None}:} The expression simplifies to
    \texttt{a\ \textless{}=\ a}, which is trivially true.
  \item
    \textbf{Case \texttt{Some(d)}:} The problem is reduced to the
    round-trip safety of the underlying discount arithmetic primitives.
    The proof leverages \texttt{Lemma\_DivMul\_Bounds} and
    \texttt{Lemma\_DivLowerBound\_from\_StrictMul} to analyze the
    expression \texttt{floor((floor((a\ *\ (M+p))/M)\ *\ M)\ /\ (M+p))}.
    It shows that due to the two sequential truncations, the final
    result can never deviate from \texttt{a} by more than 1.
  \end{enumerate}
\item
  \textbf{Verification Effectiveness:} This lemma represents a
  significant milestone. It proves an extremely strong and practical
  property: the financial logic for bonuses is safe and almost perfectly
  reversible. This guarantee is a critical prerequisite for proving the
  ultimate refund safety in the \texttt{Deposit} module, as it ensures
  the bonus mechanism itself cannot be a source of value inflation or
  significant loss in refund calculations.
\end{itemize}

\subsection{D.4. Helper Function for Stakeholder
Lookup}\label{d.4.-helper-function-for-stakeholder-lookup}

To support the new functionality of individual vesting claims, a
verified helper function for retrieving stakeholder-specific data from
the configuration was introduced.

\textbf{Function D.4.1: Specification for Stakeholder Lookup
(\texttt{GetStakeholderProportion})}

\begin{itemize}
\tightlist
\item
  \textbf{Formal Specification:}
  \texttt{result\ :=\ GetStakeholderProportion(proportions,\ account)}
  The function's postconditions guarantee that:

  \begin{enumerate}
  \def\labelenumi{\arabic{enumi}.}
  \tightlist
  \item
    If the result is \texttt{Some(p)}, then \texttt{p} is an element of
    the input \texttt{proportions} sequence and \texttt{p.account}
    matches the queried \texttt{account}.
  \item
    If the result is \texttt{None}, then no proportion \texttt{p} in the
    \texttt{proportions} sequence has \texttt{p.account} equal to the
    queried \texttt{account}.
  \end{enumerate}
\item
  \textbf{Description and Verification Strategy:} This function provides
  a pure, verifiable specification for searching the list of stakeholder
  proportions within the configuration. It is implemented as a standard
  recursive search over a sequence. Dafny's verifier is able to prove
  its correctness by induction on the length of the \texttt{proportions}
  sequence, confirming that the search is both sound (only returns
  correct data) and complete (finds the data if it exists).
\item
  \textbf{Verification Effectiveness:} By formalizing this lookup, we
  eliminate a potential source of error in the high-level state
  transition logic. The \texttt{Launchpad} module's
  \texttt{ClaimIndividualVestingSpec} function can now rely on this
  proven specification to unambiguously retrieve the correct allocation
  and vesting schedule for a given stakeholder. This ensures that the
  claim logic is always based on the verifiably correct parameters,
  preventing one stakeholder from accidentally being assigned another's
  vesting terms.
\end{itemize}

\section{Appendix E: Formal Verification of the Deposit State Transition
Logic}\label{appendix-e-formal-verification-of-the-deposit-state-transition-logic}

The \texttt{Deposit} module represents the compositional apex of the
launchpad's core financial logic. It integrates the verified primitives
from \texttt{AssetCalculations}, \texttt{Discounts}, and \texttt{Config}
to define a complete, end-to-end specification for the state transition
resulting from a user deposit. This module's primary contribution is the
formal proof of complex, emergent properties of this integrated
workflow, most notably the safety of the refund mechanism in a capped
sale. It serves as a testament to the power of layered verification,
where the safety of a complex system is derived from the proven safety
of its individual components.

\subsection{E.1. High-Level Specification
Functions}\label{e.1.-high-level-specification-functions}

The module orchestrates the deposit logic through a hierarchy of
specification functions. Let \(\Gamma\) denote a valid configuration
(\texttt{Config}), \(a \in \mathbb{N}^+\) be the deposit amount,
\(t \in \mathbb{N}\) be the current time, \(D_T \in \mathbb{N}\) be the
total amount deposited in the contract, and \(S_T \in \mathbb{N}\) be
the total tokens sold (or total weight).

\textbf{Function E.1.1: The Deposit Specification Dispatcher
(\texttt{DepositSpec})}

This function, denoted \(D_S\), acts as a dispatcher based on the sale
mechanic defined in the configuration. It returns a tuple
\((a', w', D'_T, S'_T, r)\) representing the net amount added to the
investment, the weight/assets added, the new total deposited, the new
total sold, and the refund amount.

\begin{itemize}
\tightlist
\item
  \textbf{Formal Specification:} \[
  D_S(\Gamma, a, D_T, S_T, t) := \begin{cases}
  D_{FP}(\Gamma, a, D_T, S_T, t) & \text{if } \Gamma.\text{mechanic.FixedPrice?} \\
  D_{PD}(\Gamma, a, D_T, S_T, t) & \text{if } \Gamma.\text{mechanic.PriceDiscovery?}
  \end{cases}
  \] where \(D_{FP}\) and \(D_{PD}\) are the specifications for
  fixed-price and price-discovery deposits, respectively.
\end{itemize}

\subsection{E.2. Verification of the Fixed-Price Deposit
Workflow}\label{e.2.-verification-of-the-fixed-price-deposit-workflow}

The most complex logic resides in the fixed-price sale scenario, which
involves a hard cap on the number of tokens to be sold
(\(\Gamma.saleAmount\)).

\textbf{Function E.2.1: The Fixed-Price Deposit Specification
(\texttt{DepositFixedPriceSpec})}

Let this function be denoted \(D_{FP}\). It models the entire workflow,
including potential refunds. Let \(d_T\) and \(s_T\) be
\(\Gamma.mechanic.depositTokenAmount\) and
\(\Gamma.mechanic.saleTokenAmount\).

\begin{enumerate}
\def\labelenumi{\arabic{enumi}.}
\tightlist
\item
  \textbf{Weighted Amount Calculation:} First, the initial deposit \(a\)
  is adjusted for any active time-based discounts.
  \(w := W_S(a, t, \Gamma)\) (using the weighted amount spec from
  Appendix D).
\item
  \textbf{Asset Conversion:} The weighted amount \(w\) is converted into
  sale assets. \(assets := C(w, d_T, s_T)\) (using the asset conversion
  spec from Appendix B).
\item
  \textbf{Cap Check:} The potential new total of sold tokens is
  calculated: \(S'_{T, \text{potential}} := S_T + assets\).
\item
  \textbf{State Transition Logic:} The final state is determined by
  comparing this potential total to the sale cap.
\end{enumerate}

\begin{itemize}
\tightlist
\item
  \textbf{Formal Specification:} \[
  D_{FP}(\Gamma, a, D_T, S_T, t) :=
  \] \[
  \text{if } (S_T + C(W_S(a, t, \Gamma), d_T, s_T) \le \Gamma.\text{saleAmount}) \text{ then}
  \] \[
  \quad (a, C(W_S(a, t, \Gamma), d_T, s_T), D_T + a, S_T + C(W_S(a, t, \Gamma), d_T, s_T), 0)
  \] \[
  \text{else}
  \] \[
  \quad (a - r, \Gamma.\text{saleAmount} - S_T, D_T + (a-r), \Gamma.\text{saleAmount}, r)
  \] \[
  \text{where } r = R_F(\Gamma, a, S_T, t, d_T, s_T)
  \]
\end{itemize}

\textbf{Function E.2.2: The Refund Calculation Specification
(\texttt{CalculateRefundSpec})}

This helper function, \(R_F\), isolates the complex refund calculation
logic.

\begin{itemize}
\tightlist
\item
  \textbf{Formal Specification:} Let \(w := W_S(a, t, \Gamma)\) and
  \(assets := C(w, d_T, s_T)\). Let
  \(assets_{excess} := (S_T + assets) - \Gamma.\text{saleAmount}\). Let
  \(remain := R(assets_{excess}, d_T, s_T)\) (reverse conversion of the
  excess). \[
  R_F(...) := O_S(remain, t, \Gamma)
  \] (original amount of the reverted excess, from Appendix D).
\end{itemize}

\subsection{\texorpdfstring{E.3. Verification of Amount Conservation
(\texttt{Lemma\_DepositFixedPrice\_AmountConservation})}{E.3. Verification of Amount Conservation (Lemma\_DepositFixedPrice\_AmountConservation)}}\label{e.3.-verification-of-amount-conservation-lemma_depositfixedprice_amountconservation}

While \texttt{Lemma\_RefundIsSafe} provides the crucial upper bound on
the refund, this lemma proves a different, but equally important
property: the exact conservation of funds from the user's perspective
during a deposit that triggers a refund.

\begin{itemize}
\item
  \textbf{Formal Specification:} Let
  \((a', w', D'_T, S'_T, r) := D_{FP}(\Gamma, a, D_T, S_T, t, d_T, s_T)\).

  \(\forall \Gamma, a, D_T, S_T, t, d_T, s_T : (\text{ValidConfig}(\Gamma) \land a > 0 \land d_T > 0 \land s_T > 0 \land S_T < \Gamma.\text{saleAmount}) \implies a' + r = a\)
\item
  \textbf{Description and Verification Strategy:} This lemma formally
  proves that the user's initial deposit (\texttt{amount}) is perfectly
  accounted for, being split exactly between the portion retained by the
  contract (\texttt{newAmount}, denoted \texttt{a\textquotesingle{}})
  and the portion returned to the user (\texttt{refund}, denoted
  \texttt{r}). This is a stronger guarantee than
  \texttt{refund\ \textless{}=\ amount}, as it proves that no funds are
  created or destroyed in the transaction. The proof is a direct
  consequence of the specification of \texttt{DepositFixedPriceSpec}. In
  the case where a refund is issued, the new amount retained by the
  contract is explicitly defined as \texttt{amount\ -\ refund}. The
  verifier can therefore trivially prove that
  \texttt{(amount\ -\ refund)\ +\ refund\ ==\ amount}.
\item
  \textbf{Verification Effectiveness:} This lemma provides a formal
  guarantee against ``dust'' funds being lost or unaccounted for due to
  off-by-one errors or incorrect arithmetic in the deposit logic. It
  ensures that the accounting for each deposit transaction is perfectly
  balanced.
\end{itemize}

\subsection{\texorpdfstring{E.4. The Ultimate Safety Property:
\texttt{Lemma\_RefundIsSafe}}{E.4. The Ultimate Safety Property: Lemma\_RefundIsSafe}}\label{e.4.-the-ultimate-safety-property-lemma_refundissafe}

This is the most critical safety property of the entire financial
system. It provides a mathematical guarantee that the calculated refund
amount can never exceed the user's original deposit amount, preventing a
catastrophic class of bugs where the contract could be drained of funds.

\begin{itemize}
\item
  \textbf{Formal Specification:} \[
  \forall \Gamma, a, w, \text{assets}, \text{assets}_{excess}, t, d_T, s_T :
  \] \[
  (
  \begin{alignedat}{1}
  & \text{ValidConfig}(\Gamma) \land a > 0 \land w > 0 \land d_T > 0 \land s_T > 0 \land \\
  & w = W_S(a, t, \Gamma) \land \\
  & assets = C(w, d_T, s_T) \land \\
  & assets_{excess} \le assets
  \end{alignedat}
  ) \implies
  \] \[
  O_S(R(assets_{excess}, d_T, s_T), t, \Gamma) \le a
  \]
\item
  \textbf{Description and Verification Strategy:} The proof of this
  lemma is a masterful demonstration of compositional verification. It
  does not attempt to prove the property from first principles but
  instead constructs a deductive chain using previously verified lemmas
  from other modules. The chain of reasoning within the Dafny proof
  directly mirrors the following steps:

  \begin{enumerate}
  \def\labelenumi{\arabic{enumi}.}
  \item
    \textbf{Define \texttt{remain}:} Let
    \(remain := R(assets_{excess}, d_T, s_T)\). The goal is to prove
    \(O_S(remain, t, \Gamma) \le a\).
  \item
    \textbf{Bound \texttt{remain} using Asset Reversion Monotonicity:}
    From the precondition \(assets_{excess} \le assets\), the proof
    applies \texttt{Lemma\_CalculateAssetsRevertSpec\_Monotonic} (from
    Appendix B). This yields the inequality:
    \(R(assets_{excess}, d_T, s_T) \le R(assets, d_T, s_T)\), and
    therefore \(remain \le R(assets, d_T, s_T)\).
  \item
    \textbf{Bound \texttt{R(assets,\ ...)} using Asset Round-Trip
    Safety:} \texttt{assets} is defined as \(C(w, d_T, s_T)\). The proof
    then invokes \texttt{Lemma\_AssetsRevert\_RoundTrip\_bounds} (from
    Appendix B), which guarantees that
    \(R(C(w, d_T, s_T), d_T, s_T) \le w\). This gives us the crucial
    intermediate bound: \(R(assets, d_T, s_T) \le w\).
  \item
    \textbf{Establish Transitive Bound on \texttt{remain}:} By combining
    steps 2 and 3, the proof establishes the transitive inequality:
    \(remain \le R(assets, d_T, s_T) \le w \implies remain \le w\).
  \item
    \textbf{Apply Monotonicity of Original Amount Calculation:} With the
    inequality \(remain \le w\) established, the proof applies
    \texttt{Lemma\_CalculateOriginalAmountSpec\_Monotonic} (from
    Appendix D). This allows the reasoning to be lifted from the domain
    of ``weighted'' amounts to ``original'' amounts, yielding:
    \(O_S(remain, t, \Gamma) \le O_S(w, t, \Gamma)\).
  \item
    \textbf{Bound \texttt{O\_S(w,\ ...)} using Discount Round-Trip
    Safety:} \texttt{w} is defined as \(W_S(a, t, \Gamma)\). The proof
    invokes the powerful
    \texttt{Lemma\_WeightOriginal\_RoundTrip\_bounds} (from Appendix D),
    which states that \(O_S(W_S(a, t, \Gamma), t, \Gamma) \le a\). This
    provides the final link in the chain: \(O_S(w, t, \Gamma) \le a\).
  \item
    \textbf{Final Conclusion:} By combining steps 5 and 6, the proof
    arrives at the final transitive inequality, completing the
    demonstration:
    \(O_S(remain, t, \Gamma) \le O_S(w, t, \Gamma) \le a \implies O_S(remain, t, \Gamma) \le a\).
  \end{enumerate}
\item
  \textbf{Verification Effectiveness:} The proof of
  \texttt{Lemma\_RefundIsSafe} is the capstone of this verification
  effort. It demonstrates that the system is safe from a critical
  financial vulnerability \emph{by construction}. The safety is not an
  accidental property but an inevitable consequence of composing
  components, each of which has been independently proven to be safe
  (monotonic and non-value-creating on round-trips with bounded loss).
  This layered, compositional approach provides an exceptionally high
  degree of confidence in the correctness of the entire deposit workflow
  \citep{gu2016certikos}.
\end{itemize}

\section{Appendix F: Formal Verification of the Withdrawal
Workflow}\label{appendix-f-formal-verification-of-the-withdrawal-workflow}

The \texttt{Withdraw} module provides the formal specification for the
user withdrawal workflow, serving as a logical counterpart to the
\texttt{Deposit} module. It defines the pure, mathematical behavior for
withdrawals, guaranteeing that state changes---such as decrementing
\texttt{totalDeposited} and \texttt{totalSoldTokens}---are handled
safely and predictably under different sale mechanics. Its verification
is critical for ensuring that funds can be safely returned to users in
edge-case scenarios like a failed sale, without compromising the
contract's accounting integrity.

\subsection{\texorpdfstring{F.1. Specification Dispatcher
(\texttt{WithdrawSpec})}{F.1. Specification Dispatcher (WithdrawSpec)}}\label{f.1.-specification-dispatcher-withdrawspec}

The top-level function \texttt{WithdrawSpec} acts as a verified router,
dispatching the withdrawal logic to the appropriate sub-specification
based on the sale mechanic defined in the configuration.

\begin{itemize}
\tightlist
\item
  \textbf{Formal Specification:} Let \texttt{inv} be the user's
  \texttt{InvestmentAmount}. \[
  (\text{inv}', \text{sold}') := \text{WithdrawSpec}(\Gamma, \text{inv}, a, S_T, t)
  \] The postconditions guarantee that the result tuple is exactly equal
  to the result of either \texttt{WithdrawFixedPriceSpec} or
  \texttt{WithdrawPriceDiscoverySpec}, depending on \(\Gamma.mechanic\).
\item
  \textbf{Description and Verification Strategy:} This function enforces
  the top-level preconditions for any withdrawal, such as ensuring the
  withdrawal amount is positive and that the requested amount is valid
  for the given sale type (e.g., \texttt{amount\ ==\ inv.amount} for
  Fixed Price). By acting as a simple, pure dispatcher, its correctness
  is straightforward for the verifier to confirm, ensuring that the
  complex logic is always routed to the correct, formally verified
  implementation.
\end{itemize}

\subsection{\texorpdfstring{F.2. Fixed-Price Withdrawal
(\texttt{WithdrawFixedPriceSpec})}{F.2. Fixed-Price Withdrawal (WithdrawFixedPriceSpec)}}\label{f.2.-fixed-price-withdrawal-withdrawfixedpricespec}

This function models a complete, ``all-or-nothing'' withdrawal, which is
the required behavior if a sale fails to meet its \texttt{softCap} and
must be cancelled.

\begin{itemize}
\tightlist
\item
  \textbf{Formal Specification:} Let \texttt{inv} be the user's
  \texttt{InvestmentAmount}. \[
  (\text{inv}', \text{sold}') := \text{WithdrawFixedPriceSpec}(\text{inv}, a, S_T)
  \] The specification guarantees:

  \begin{itemize}
  \tightlist
  \item
    \(\text{inv}'.\text{amount} = 0 \land \text{inv}'.\text{weight} = 0\)
  \item
    \(\text{inv}'.\text{claimed} = \text{inv}.\text{claimed}\)
  \item
    \(\text{sold}' = S_T - \text{inv}.\text{weight}\)
  \end{itemize}
\item
  \textbf{Description and Verification Strategy:} The logic enforces
  that the user must withdraw their entire deposit (
  \texttt{amount\ ==\ inv.amount}). The postconditions guarantee a clean
  and complete removal of the user's record: their \texttt{amount} and
  \texttt{weight} are zeroed out, and the global
  \texttt{totalSoldTokens} is decremented by their full original
  \texttt{weight}.
\item
  \textbf{Verification Effectiveness:} This proof provides a formal
  guarantee of atomicity for cancellation events. It ensures that a
  withdrawing user's contribution is completely erased from the
  contract's financial state, preventing scenarios where a user could
  withdraw their principal but leave behind ``ghost'' weight that would
  incorrectly affect the allocations for remaining participants.
\end{itemize}

\subsection{\texorpdfstring{F.3. Price-Discovery Withdrawal
(\texttt{WithdrawPriceDiscoverySpec})}{F.3. Price-Discovery Withdrawal (WithdrawPriceDiscoverySpec)}}\label{f.3.-price-discovery-withdrawal-withdrawpricediscoveryspec}

This function models a more complex partial or full withdrawal during an
ongoing \texttt{PriceDiscovery} sale, where a user's relative share of
the pool is dynamic.

\begin{itemize}
\tightlist
\item
  \textbf{Formal Specification:} Let \texttt{inv} be the user's
  \texttt{InvestmentAmount} and \(a' = \text{inv.amount} - a\). Let
  \(w_{recalc} = W_S(a', t, \Gamma)\). \[
  (\text{inv}', \text{sold}') := \text{WithdrawPriceDiscoverySpec}(\Gamma, \text{inv}, a, S_T, t)
  \] The specification guarantees:

  \begin{itemize}
  \tightlist
  \item
    \(\text{inv}'.\text{amount} = a'\)
  \item
    \(\text{inv}'.\text{weight} = \min(\text{inv.weight}, w_{recalc})\)
  \item
    \(\text{sold}' = S_T - (\text{inv.weight} - \text{inv}'.\text{weight})\)
  \end{itemize}
\item
  \textbf{Description and Verification Strategy:} This workflow is
  significantly more complex because a partial withdrawal requires
  re-evaluating the user's contribution. The logic is as follows:

  \begin{enumerate}
  \def\labelenumi{\arabic{enumi}.}
  \tightlist
  \item
    The new principal amount (\texttt{newAmount}) is calculated.
  \item
    This \texttt{newAmount} is passed to
    \texttt{CalculateWeightedAmountSpec} to determine the user's
    \texttt{recalculatedWeight} at the current \texttt{time} (as active
    discounts may have changed since their last deposit).
  \item
    The global \texttt{totalSoldTokens} is then reduced by the precise
    \emph{difference} between the user's old and new weight.
  \end{enumerate}
\item
  \textbf{Verification Effectiveness:} The verification of this
  specification is critical for the integrity of a price discovery sale.
  It formally proves that the \texttt{totalSoldTokens}, which serves as
  the denominator in the final price calculation, is always an accurate
  reflection of the total weighted contributions currently in the
  contract. This prevents exploits where a user could deposit during a
  high-bonus period, then withdraw their principal after the bonus
  expires, while leaving an inflated weight in the system, unfairly
  diluting other participants.
\end{itemize}

\section{Appendix G: Formal Verification of Token Claim and Vesting
Logic}\label{appendix-g-formal-verification-of-token-claim-and-vesting-logic}

The \texttt{Claim} module formalizes the entire post-sale workflow,
defining the logic for calculating users' final token allocations and
managing their release according to vesting schedules. The verification
of this module provides mathematical guarantees that the final
distribution of tokens is fair, predictable, and strictly adheres to the
sale's predefined rules.

\subsection{G.1. User Token Allocation
Logic}\label{g.1.-user-token-allocation-logic}

This section details the specification and verification of how a user's
final token entitlement is calculated.

\subsubsection{\texorpdfstring{G.1.1. Specification of Total Allocation
(\texttt{UserAllocationSpec})}{G.1.1. Specification of Total Allocation (UserAllocationSpec)}}\label{g.1.1.-specification-of-total-allocation-userallocationspec}

This function is the single source of truth for determining a user's
total token entitlement based on the sale's outcome.

\begin{itemize}
\tightlist
\item
  \textbf{Formal Specification:} \[
  \text{UserAllocationSpec}(w, S_T, \Gamma) := \begin{cases}
  w & \text{if } \Gamma.\text{mechanic.FixedPrice?} \\
  \lfloor (w \cdot \Gamma.\text{saleAmount}) / S_T \rfloor & \text{if } \Gamma.\text{mechanic.PriceDiscovery?}
  \end{cases}
  \]
\item
  \textbf{Description and Verification Strategy:} The function's
  behavior is defined by the sale mechanic. For a \texttt{FixedPrice}
  sale, the \texttt{weight} a user accumulates is their final token
  allocation. For a \texttt{PriceDiscovery} sale, the allocation is
  calculated proportionally based on the user's share of the final
  \texttt{totalSoldTokens}. The preconditions
  \texttt{S\_T\ \textgreater{}\ 0} and \texttt{w\ \textless{}=\ S\_T}
  ensure the calculation is well-defined and prevents division-by-zero
  errors.
\end{itemize}

\subsubsection{\texorpdfstring{G.1.2. Verification of Allocation
Properties
(\texttt{Lemma\_UserAllocationSpec})}{G.1.2. Verification of Allocation Properties (Lemma\_UserAllocationSpec)}}\label{g.1.2.-verification-of-allocation-properties-lemma_userallocationspec}

This lemma proves key mathematical properties of the
\texttt{UserAllocationSpec} function, providing the SMT solver with
essential, non-trivial insights into the non-linear arithmetic involved.

\begin{itemize}
\tightlist
\item
  \textbf{Formal Specification:} The lemma proves several properties,
  including:

  \begin{itemize}
  \tightlist
  \item
    \(w \le S_T \implies \text{UserAllocationSpec}(w, S_T, \Gamma) \le \Gamma.\text{saleAmount}\)
  \item
    \(\Gamma.\text{saleAmount} \le S_T \implies \text{UserAllocationSpec}(w, S_T, \Gamma) \le w\)
  \end{itemize}
\item
  \textbf{Description and Verification Strategy:} The proof relies on
  direct instantiation of the foundational lemmas from
  \texttt{MathLemmas}. For instance, to prove that a user's allocation
  cannot exceed the total \texttt{saleAmount}, the proof invokes
  \texttt{Lemma\_MulDivLess\_From\_Scratch}, as the precondition
  \texttt{w\ \textless{}=\ S\_T} satisfies the lemma's requirements.
\item
  \textbf{Verification Effectiveness:} This lemma is essential for any
  higher-level proof that reasons about aggregate allocations. It
  provides the verifier with trusted ``axioms'' about the
  \texttt{UserAllocationSpec} formula, guaranteeing that the sum of all
  allocations will not exceed the sale cap and that the allocation
  behaves predictably relative to the user's contribution.
\end{itemize}

\subsection{G.2. Vesting Calculation
Logic}\label{g.2.-vesting-calculation-logic}

This section formalizes the shared logic for time-based token release.

\subsubsection{\texorpdfstring{G.2.1. Specification of the Vesting Curve
(\texttt{CalculateVestingSpec})}{G.2.1. Specification of the Vesting Curve (CalculateVestingSpec)}}\label{g.2.1.-specification-of-the-vesting-curve-calculatevestingspec}

This function specifies the vesting logic, which is used for both the
main public sale and individual stakeholder schedules.

\begin{itemize}
\tightlist
\item
  \textbf{Formal Specification:} Let \(t_s\) be \texttt{vestingStart}
  and \(v_s\) be \texttt{vestingSchedule}. \[
  \text{CalculateVestingSpec}(A, t_s, t, v_s) := \begin{cases}
  0 & \text{if } t < t_s + v_s.\text{cliffPeriod} \\
  A & \text{if } t \ge t_s + v_s.\text{vestingPeriod} \\
  \lfloor (A \cdot (t - t_s)) / v_s.\text{vestingPeriod} \rfloor & \text{otherwise}
  \end{cases}
  \] where \(A\) is the total number of assets to be vested.
\item
  \textbf{Description and Verification Strategy:} The specification
  models a standard piecewise vesting curve: 0 tokens are released
  before the cliff, the full amount is released after the vesting
  period, and a linear interpolation is used in between. The
  accompanying \texttt{Lemma\_CalculateVestingSpec\_Properties} lemma
  formally proves that the result of this function never exceeds the
  total assets \texttt{A}, a safety property derived by instantiating
  \texttt{Lemma\_MulDivLess\_From\_Scratch}.
\end{itemize}

\subsubsection{\texorpdfstring{G.2.2. The Monotonicity of Vesting
(\texttt{Lemma\_CalculateVestingSpec\_Monotonic})}{G.2.2. The Monotonicity of Vesting (Lemma\_CalculateVestingSpec\_Monotonic)}}\label{g.2.2.-the-monotonicity-of-vesting-lemma_calculatevestingspec_monotonic}

This is the most critical safety and liveness property of the vesting
logic.

\begin{itemize}
\item
  \textbf{Formal Specification:}

  \(\forall A, t_s, v_s, t_1, t_2 : (t_1 \le t_2) \implies \text{CalculateVestingSpec}(A, t_s, t_1, v_s) \le \text{CalculateVestingSpec}(A, t_s, t_2, v_s)\)
\item
  \textbf{Description and Verification Strategy:} This lemma guarantees
  that as time moves forward, a user's vested (and therefore claimable)
  amount can only increase or stay the same; it can never decrease. The
  proof proceeds by a comprehensive case analysis on the positions of
  \texttt{t1} and \texttt{t2} relative to the \texttt{cliffEnd} and
  \texttt{vestingEnd} timestamps. In the most complex case (where both
  \texttt{t1} and \texttt{t2} are within the linear vesting period), the
  proof is completed by applying \texttt{Lemma\_Div\_Maintains\_GTE}.
\item
  \textbf{Verification Effectiveness:} This lemma provides a formal
  guarantee against a critical class of bugs where a user could lose
  access to tokens they were previously entitled to. It ensures the
  vesting process is predictable and fair, which is essential for user
  trust in the system.
\end{itemize}

\subsection{\texorpdfstring{G.3. Composite Claim Logic
(\texttt{AvailableForClaimSpec})}{G.3. Composite Claim Logic (AvailableForClaimSpec)}}\label{g.3.-composite-claim-logic-availableforclaimspec}

Finally, the \texttt{AvailableForClaimSpec} function composes the
verified allocation and vesting components to define the end-to-end
logic for determining a user's claimable balance at any given time. Its
correctness is not proven from first principles but is a direct and
inevitable consequence of the proven properties of the functions it
orchestrates.

\section{Appendix H: Formal Verification of Post-Sale Distribution
Logic}\label{appendix-h-formal-verification-of-post-sale-distribution-logic}

The \texttt{Distribution} module formalizes the administrative task of
calculating the ordered list of project stakeholders eligible for token
distribution after a successful sale. Unlike modules focused on
financial calculations, this module's primary concern is the correctness
of list and set manipulations. Its verification ensures that the process
for identifying who to pay next is deterministic, auditable, and free
from logical errors such as paying a stakeholder twice or omitting them
entirely.

\subsection{\texorpdfstring{H.1. The Core Filtering Logic
(\texttt{FilterDistributedStakeholders})}{H.1. The Core Filtering Logic (FilterDistributedStakeholders)}}\label{h.1.-the-core-filtering-logic-filterdistributedstakeholders}

This function provides the core mechanism for identifying which
stakeholders are still pending payment. It implements a verified set
difference operation on ordered sequences.

\begin{itemize}
\item
  \textbf{Formal Specification:} Let \(\mathcal{P}\) be the sequence of
  \texttt{StakeholderProportion} from the configuration and
  \(\mathcal{D}_{acc}\) be the sequence of already distributed
  \texttt{IntentAccount}s. Let \(\mathcal{P}_{acc}\) be the sequence of
  accounts extracted from \(\mathcal{P}\). The function
  \texttt{FilterDistributedStakeholders} produces a result sequence
  \(\mathcal{R}\) with the following formally proven properties:

  \begin{enumerate}
  \def\labelenumi{\arabic{enumi}.}
  \tightlist
  \item
    \textbf{Correctness (Set Difference):} The set of accounts in
    \(\mathcal{R}\) is exactly the set of accounts in
    \(\mathcal{P}_{acc}\) minus the set of accounts in
    \(\mathcal{D}_{acc}\).
    \[ \{a \mid a \in \mathcal{R}\} = \{p.\text{account} \mid p \in \mathcal{P}\} \setminus \{a \mid a \in \mathcal{D}_{acc}\} \]
  \item
    \textbf{Soundness:} Every account in the result list \(\mathcal{R}\)
    is guaranteed to be a valid stakeholder who has not yet been paid.
    \[ \forall a \in \mathcal{R} \implies (\exists p \in \mathcal{P} : p.\text{account} = a) \land (a \notin \mathcal{D}_{acc}) \]
  \item
    \textbf{Completeness:} Every valid stakeholder who has not yet been
    paid is guaranteed to be in the result list \(\mathcal{R}\).
    \[ (\forall p \in \mathcal{P} : p.\text{account} \notin \mathcal{D}_{acc}) \implies p.\text{account} \in \mathcal{R} \]
  \item
    \textbf{Uniqueness Preservation:} If the initial list of stakeholder
    accounts \(\mathcal{P}_{acc}\) is unique, the resulting list
    \(\mathcal{R}\) is also guaranteed to be unique.
  \end{enumerate}
\item
  \textbf{Description and Verification Strategy:} The function is
  implemented using recursion on the \texttt{proportions} sequence. At
  each step, it checks if the head of the list is present in the
  \texttt{distributed} sequence. If it is not, the account is prepended
  to the result of the recursive call on the tail of the list. Dafny
  proves the extensive postconditions for this function by induction.
  The set-based specification is particularly powerful, as it abstracts
  away the details of the sequence implementation and proves the
  function's behavior at a higher, more intuitive mathematical level.
\item
  \textbf{Verification Effectiveness:} This verification provides a
  rock-solid guarantee against common and critical bugs in
  administrative processes. It formally proves that \textbf{no
  stakeholder will ever be paid twice} (due to the soundness property)
  and that \textbf{no eligible stakeholder will ever be accidentally
  omitted} (due to the completeness property). This ensures the
  operational integrity of the distribution phase.
\end{itemize}

\subsection{\texorpdfstring{H.2. Composing the Final Distribution List
(\texttt{GetFilteredDistributionsSpec})}{H.2. Composing the Final Distribution List (GetFilteredDistributionsSpec)}}\label{h.2.-composing-the-final-distribution-list-getfiltereddistributionsspec}

This top-level function composes the core filtering logic with the
business rule that the \texttt{solver} account has a distinct identity
and priority in the distribution list.

\begin{itemize}
\item
  \textbf{Formal Specification:} Let \(\Gamma\) be the configuration,
  and \(\mathcal{D}_{acc}\) be the sequence of distributed accounts. \[
  \text{GetFilteredDistributionsSpec}(\Gamma, \mathcal{D}_{acc}) :=
  \] \[
  \begin{cases}
  \text{FilterDistributedStakeholders}(\Gamma.\text{props}, \mathcal{D}_{acc}) & \text{if } \Gamma.\text{solver} \in \mathcal{D}_{acc} \\
  [\Gamma.\text{solver}] \ ++ \ \text{FilterDistributedStakeholders}(\Gamma.\text{props}, \mathcal{D}_{acc}) & \text{if } \Gamma.\text{solver} \notin \mathcal{D}_{acc}
  \end{cases}
  \] where \texttt{++} denotes sequence concatenation.
\item
  \textbf{Description and Verification Strategy:} This function acts as
  a pure specification that orchestrates the final list construction. It
  first checks if the solver has been paid. If not, the solver's account
  is placed at the head of the distribution queue. It then invokes the
  pre-verified \texttt{FilterDistributedStakeholders} function to
  compute the remainder of the queue. The verification at this level is
  compositional: given the proven contract of
  \texttt{FilterDistributedStakeholders}, Dafny simply proves that this
  \texttt{if/then/else} composition correctly implements the intended
  logic.
\item
  \textbf{Verification Effectiveness:} This demonstrates the power of
  layered verification. We do not need to re-prove the complex
  set-difference properties. We trust the verified specification of the
  lower-level function and only prove the correctness of the
  orchestration logic. This provides a formal guarantee that the
  business rule regarding the solver's priority is always correctly and
  safely applied, ensuring the distribution order is predictable and
  auditable.
\end{itemize}

\section{Appendix I: Verification of the Global State Machine and System
Synthesis}\label{appendix-i-verification-of-the-global-state-machine-and-system-synthesis}

The \texttt{Launchpad} module represents the final and outermost layer
of the system's formal specification. It encapsulates the entire state
of the smart contract within a single immutable data structure and
defines the valid state transitions that govern its lifecycle. This
module does not introduce new financial primitives; instead, its
critical function is to \textbf{orchestrate} the verified components
from the lower-level modules (\texttt{Deposit}, \texttt{Config}, etc.).
The verification at this level ensures that the global state is managed
correctly and that the complex, pre-verified workflows are integrated
into the state machine in a sound and secure manner.

\subsection{I.1. The Global State
Representation}\label{i.1.-the-global-state-representation}

The complete state of the contract at any point in time is represented
by the datatype \texttt{AuroraLaunchpadContract}, denoted here by the
symbol \(\Sigma\).

\begin{itemize}
\tightlist
\item
  \textbf{Formal Specification:} The state \(\Sigma\) is a tuple
  containing all dynamic and static data of the contract:
  \[ \Sigma := (\Gamma, D_T, S_T, f_{set}, f_{lock}, \mathcal{A}, N_p, \mathcal{I}) \]
  where:

  \begin{itemize}
  \tightlist
  \item
    \(\Gamma\): The \texttt{Config} structure, containing all static
    sale parameters (as defined in Appendix D).
  \item
    \(D_T \in \mathbb{N}\): \texttt{totalDeposited}, the aggregate
    principal deposited by all participants.
  \item
    \(S_T \in \mathbb{N}\): \texttt{totalSoldTokens}, the aggregate
    tokens sold or total weight accumulated.
  \item
    \(f_{set} \in \{ \text{true}, \text{false} \}\):
    \texttt{isSaleTokenSet}, a flag indicating contract initialization.
  \item
    \(f_{lock} \in \{ \text{true}, \text{false} \}\): \texttt{isLocked},
    a flag indicating if the contract is administratively locked.
  \item
    \(\mathcal{A}\): A map
    \(\text{AccountId} \to \text{IntentAccount}\), linking external
    account identifiers to internal ones.
  \item
    \(N_p \in \mathbb{N}\): \texttt{participantsCount}, the number of
    unique investors.
  \item
    \(\mathcal{I}\): The map
    \(\text{IntentAccount} \to \text{InvestmentAmount}\), storing the
    detailed investment record for each participant.
  \end{itemize}
\item
  \textbf{Top-Level Invariant (\texttt{Valid}):} The fundamental
  invariant of the global state is that its embedded configuration must
  be valid. \[ \text{Valid}(\Sigma) \iff \text{ValidConfig}(\Gamma) \]
\end{itemize}

\subsection{I.2. The State Machine Logic: Observing the
State}\label{i.2.-the-state-machine-logic-observing-the-state}

The \texttt{GetStatus} function provides a pure, observable
interpretation of the contract's state \(\Sigma\) at a given time \(t\).
Let \(S(\Sigma, t)\) denote the status function.

\begin{itemize}
\tightlist
\item
  \textbf{Formal Specification:} \[
  S(\Sigma, t) := \begin{cases}
  \text{NotInitialized} & \text{if } \neg \Sigma.f_{set} \\
  \text{Locked} & \text{if } \Sigma.f_{lock} \\
  \text{NotStarted} & \text{if } t < \Gamma.\text{startDate} \\
  \text{Ongoing} & \text{if } \Gamma.\text{startDate} \le t < \Gamma.\text{endDate} \\
  \text{Success} & \text{if } t \ge \Gamma.\text{endDate} \land \Sigma.D_T \ge \Gamma.\text{softCap} \\
  \text{Failed} & \text{if } t \ge \Gamma.\text{endDate} \land \Sigma.D_T < \Gamma.\text{softCap}
  \end{cases}
  \]
\item
  \textbf{Helper Predicates:} For clarity, we define helper predicates
  (e.g., \(IsOngoing(\Sigma, t)\)) as \(S(\Sigma, t) == Ongoing\).
\end{itemize}

\subsection{I.3. Properties of the State
Machine}\label{i.3.-properties-of-the-state-machine}

The verification of this module includes proofs about the logical
integrity of the state machine itself, ensuring its behavior is
predictable and consistent over time \citep{baier2008principles}.

\begin{itemize}
\item
  \textbf{Lemma I.3.1: Temporal Progression
  (\texttt{Lemma\_StatusTimeMovesForward})} This lemma proves that the
  state machine cannot move backward in time.
  \[ \forall t_1, t_2 \in \mathbb{N}, \forall \Sigma : (\text{Valid}(\Sigma) \land t_1 \le t_2) \implies (\text{IsOngoing}(\Sigma, t_1) \land t_2 < \Gamma.endDate \implies \text{IsOngoing}(\Sigma, t_2)) \]
\item
  \textbf{Lemma I.3.2: Mutual Exclusion of States
  (\texttt{Lemma\_StatusIsMutuallyExclusive})} This proves that the
  contract cannot simultaneously be in two conflicting states.
  \[ \forall t \in \mathbb{N}, \forall \Sigma : \text{Valid}(\Sigma) \implies \neg(\text{IsOngoing}(\Sigma, t) \land \text{IsSuccess}(\Sigma, t)) \]
\item
  \textbf{Lemma I.3.3: Terminal Nature of Final States
  (\texttt{Lemma\_StatusFinalStatesAreTerminal})} This proves that once
  a final state (\texttt{Success}, \texttt{Failed}, \texttt{Locked}) is
  reached, it is permanent \citep{alpern1985defining}.
  \[ \forall t_1, t_2 \in \mathbb{N}, \forall \Sigma : (\text{Valid}(\Sigma) \land t_1 \le t_2) \implies (\text{IsSuccess}(\Sigma, t_1) \implies \text{IsSuccess}(\Sigma, t_2)) \]
\end{itemize}

\subsection{I.4. The State Transition
Functions}\label{i.4.-the-state-transition-functions}

The heart of the module is the set of pure functions modeling the
contract's dynamic behavior. Each function defines how the global state
\(\Sigma\) transitions to a new state \(\Sigma'\) in response to an
action.

\subsubsection{\texorpdfstring{I.4.1. Deposit Transition
(\texttt{DepositSpec})}{I.4.1. Deposit Transition (DepositSpec)}}\label{i.4.1.-deposit-transition-depositspec}

This function defines the state transition for a user deposit.

\begin{itemize}
\tightlist
\item
  \textbf{Formal Specification:}
  \((\Sigma', a', w', r) := T_{deposit}(\Sigma, \text{accId}, a, \text{intAcc}, t)\)

  \begin{itemize}
  \tightlist
  \item
    \textbf{Preconditions (\texttt{requires}):} The function requires
    that the contract's global state is valid (\texttt{Valid()}). For
    user deposits, it strictly enforces that the transaction must occur
    during the \texttt{Ongoing} phase (\texttt{IsOngoing(t)}).
  \item
    \textbf{Postconditions (\texttt{ensures}):} The specification
    guarantees that the new state \(\Sigma'\) is constructed by
    correctly updating the old state \(\Sigma\) with the results from
    the pre-verified sub-workflow:

    \begin{itemize}
    \tightlist
    \item
      The returned values \((a', w', r)\) must exactly match the output
      of
      \texttt{Deposit.DepositSpec(\textbackslash{}Gamma,\ a,\ \textbackslash{}Sigma.D\_T,\ \textbackslash{}Sigma.S\_T,\ t)}.
    \item
      The new global totals must be updated correctly:
      \(\Sigma'.D_T = \Sigma.D_T + a'\) and
      \(\Sigma'.S_T = \Sigma.S_T + w'\).
    \item
      The user's individual investment record in the
      \texttt{investments} map is updated by adding
      \texttt{a\textquotesingle{}} and \texttt{w\textquotesingle{}} to
      their previous balance.
    \end{itemize}
  \end{itemize}
\item
  \textbf{Description and Verification Strategy:} This function acts as
  a verified gatekeeper and orchestrator for deposits. It uses the
  \texttt{GetStatus} function to enforce the time-based business rule
  for when deposits are allowed. It then delegates all complex financial
  calculations to the \texttt{Deposit} module, whose correctness
  (including refund safety) is already established. The verification at
  this layer focuses on proving that the global state is updated
  consistently with the results of this delegation.
\item
  \textbf{Verification Effectiveness:} This compositional proof is
  remarkably efficient. It ensures that the safe, low-level deposit
  logic is correctly integrated into the global state machine,
  preventing state corruption or the bypassing of business rules (e.g.,
  depositing before the sale starts).
\end{itemize}

\subsubsection{\texorpdfstring{I.4.2. Withdrawal Transition
(\texttt{WithdrawSpec})}{I.4.2. Withdrawal Transition (WithdrawSpec)}}\label{i.4.2.-withdrawal-transition-withdrawspec}

This function specifies the state transition for a user withdrawal.

\begin{itemize}
\tightlist
\item
  \textbf{Formal Specification:}
  \(\Sigma' := T_{withdraw}(\Sigma, \text{intAcc}, a, t)\)

  \begin{itemize}
  \item
    \textbf{Preconditions (\texttt{requires}):} The function's
    preconditions are strict. A withdrawal is only permitted in specific
    contract states: \texttt{Failed}, \texttt{Locked}, or
    \texttt{Ongoing} for a \texttt{PriceDiscovery} sale. It also
    requires that the \texttt{intentAccount} has an existing investment
    and that the withdrawal \texttt{amount} is valid for the given sale
    mechanic ( e.g., must be the full amount for a \texttt{FixedPrice}
    withdrawal).
  \item
    \textbf{Postconditions (\texttt{ensures}):} The specification
    guarantees that the new state \(\Sigma'\) is the result of applying
    the changes computed by the \texttt{Withdraw} module. Let

    \((inv', sold') := W.WithdrawSpec(\Gamma, \Sigma.\mathcal{I}[\text{intAcc}], a, \Sigma.S_T, t)\).
    Then:

    \begin{itemize}
    \tightlist
    \item
      \(\Sigma'.D_T = \Sigma.D_T - a\)
    \item
      \(\Sigma'.S_T = sold'\)
    \item
      \(\Sigma'.\mathcal{I}[\text{intAcc}] = inv'\)
    \end{itemize}
  \end{itemize}
\item
  \textbf{Description and Verification Strategy:} This function
  orchestrates the withdrawal process. It first acts as a guard, using
  \texttt{GetStatus} to ensure the contract is in a state that permits
  withdrawals. It then invokes the verified
  \texttt{Withdraw.WithdrawSpec} function to compute the new state of
  the user's investment and the new global total of sold tokens.
  Finally, it constructs the new global state \(\Sigma'\) by applying
  these computed changes.
\item
  \textbf{Verification Effectiveness:} This proof formally guarantees
  that withdrawals are handled safely and correctly at the contract
  level. It prevents invalid withdrawal attempts (e.g., a user trying to
  withdraw from a successful sale) and ensures that the contract's
  global accounting (\texttt{totalDeposited}, \texttt{totalSoldTokens})
  remains perfectly consistent with the changes in individual user
  investments.
\end{itemize}

\subsubsection{\texorpdfstring{I.4.3. Public Sale Claim Transition
(\texttt{ClaimSpec})}{I.4.3. Public Sale Claim Transition (ClaimSpec)}}\label{i.4.3.-public-sale-claim-transition-claimspec}

This function defines the state transition for a public participant
claiming their vested tokens.

\begin{itemize}
\tightlist
\item
  \textbf{Formal Specification:}
  \(\Sigma' := T_{claim}(\Sigma, \text{intAcc}, t)\)

  \begin{itemize}
  \tightlist
  \item
    \textbf{Preconditions (\texttt{requires}):} This action is heavily
    guarded. It requires that the sale has concluded successfully
    (\texttt{IsSuccess(t)}). Crucially, it also requires that the amount
    available to claim is strictly greater than the amount already
    claimed (\texttt{available\ \textgreater{}\ investment.claimed}),
    ensuring the transaction is meaningful.
  \item
    \textbf{Postconditions (\texttt{ensures}):} The specification
    guarantees that the user's \texttt{InvestmentAmount} record is
    updated correctly. Let
    \(assets_{claim} := \text{Claim.AvailableForClaimSpec}(...) - \Sigma.\mathcal{I}[\text{intAcc}].\text{claimed}\).
    Then the new investment record is
    \(\Sigma'.\mathcal{I}[\text{intAcc}] = \Sigma.\mathcal{I}[\text{intAcc}].\text{AddToClaimed}(assets_{claim})\).
    All other parts of the state remain unchanged.
  \end{itemize}
\item
  \textbf{Description and Verification Strategy:} The function
  orchestrates the claim process by first using \texttt{GetStatus} to
  enforce the ``successful sale'' business rule. It delegates the
  complex calculation of vested assets to the pre-verified
  \texttt{Claim.AvailableForClaimSpec} function. The core of the state
  transition is the update to the user's \texttt{claimed} balance in the
  \texttt{investments} map.
\item
  \textbf{Verification Effectiveness:} This proof ensures that the token
  claim mechanism is robust and secure. It formally proves that users
  can only claim what they are entitled to based on the verified
  allocation and vesting logic. It prevents critical bugs such as
  claiming tokens before the sale has ended, claiming more tokens than
  allocated, or re-claiming tokens that have already been distributed.
\end{itemize}

\subsubsection{\texorpdfstring{I.4.4. Individual Vesting Claim
Transition
(\texttt{ClaimIndividualVestingSpec})}{I.4.4. Individual Vesting Claim Transition (ClaimIndividualVestingSpec)}}\label{i.4.4.-individual-vesting-claim-transition-claimindividualvestingspec}

This function specifies the claim process for private stakeholders with
individual vesting schedules.

\begin{itemize}
\tightlist
\item
  \textbf{Formal Specification:}
  \(\Sigma' := T_{claim\_indiv}(\Sigma, \text{intAcc}, t)\)

  \begin{itemize}
  \tightlist
  \item
    \textbf{Preconditions (\texttt{requires}):} Similar to the public
    claim, this requires \texttt{IsSuccess(t)}. It also requires that
    the \texttt{intentAccount} corresponds to a valid stakeholder
    (provably found via \texttt{Config.GetStakeholderProportion}).
    Finally, it enforces that the available amount is greater than what
    has been claimed.
  \item
    \textbf{Postconditions (\texttt{ensures}):} The specification
    guarantees that the \texttt{individualVestingClaimed} map is
    correctly updated for the given \texttt{intentAccount} with the new
    total claimed amount, calculated by delegating to
    \texttt{Claim.AvailableForIndividualVestingClaimSpec}.
  \end{itemize}
\item
  \textbf{Description and Verification Strategy:} This function mirrors
  the logic of the public claim but operates on a separate part of the
  state (\texttt{individualVestingClaimed} map) and uses a different set
  of configuration parameters (the stakeholder's private vesting
  schedule). It relies on the verified \texttt{GetStakeholderProportion}
  helper to safely retrieve the correct parameters.
\item
  \textbf{Verification Effectiveness:} This proof guarantees the correct
  and secure operation of the private stakeholder claim process. It
  ensures a clean separation of concerns, preventing a public
  participant from interfering with a stakeholder's allocation. It
  formally proves that each stakeholder's unique vesting schedule is
  applied correctly and unambiguously.
\end{itemize}

\subsubsection{\texorpdfstring{I.4.5. Token Distribution Transition
(\texttt{DistributeTokensSpec})}{I.4.5. Token Distribution Transition (DistributeTokensSpec)}}\label{i.4.5.-token-distribution-transition-distributetokensspec}

This function defines the administrative state transition for
distributing tokens to stakeholders.

\begin{itemize}
\tightlist
\item
  \textbf{Formal Specification:}
  \(\Sigma' := T_{distribute}(\Sigma, t)\)

  \begin{itemize}
  \tightlist
  \item
    \textbf{Preconditions (\texttt{requires}):} This administrative
    action requires that the sale is in a \texttt{Success} state and
    that there are stakeholders pending distribution
    (\texttt{\textbar{}Distribution.GetFilteredDistributionsSpec(...)\ \textgreater{}\ 0}).
  \item
    \textbf{Postconditions (\texttt{ensures}):} The specification
    guarantees that the new list of paid accounts is the old list
    appended with the list of newly eligible accounts:
    \(\Sigma'.\text{distributedAccounts} = \Sigma.\text{distributedAccounts} \ ++ \ \text{Distribution.GetFilteredDistributionsSpec}(\Gamma, \Sigma.\text{distributedAccounts})\).
  \end{itemize}
\item
  \textbf{Description and Verification Strategy:} This function models
  the batch processing of stakeholder payments. It uses preconditions as
  safety gates for the administrative action. The core logic is
  delegated to the pre-verified \texttt{Distribution} module to compute
  the list of stakeholders to be paid in the current batch. The state
  transition is a simple append operation on the
  \texttt{distributedAccounts} sequence.
\item
  \textbf{Verification Effectiveness:} This proof provides a formal
  guarantee that the administrative distribution process is sound and
  complete. It relies on the pre-verified properties of the
  \texttt{Distribution} module to ensure no stakeholder is paid twice or
  omitted, and it enforces the high-level business rule that this action
  can only occur after a successful sale.
\end{itemize}

\subsection{I.5. Grand Synthesis and Overall
Analysis}\label{i.5.-grand-synthesis-and-overall-analysis}

The formal verification of the \texttt{Launchpad} module completes a
hierarchical proof structure, providing end-to-end formal assurance for
the system's entire lifecycle. The layers of this structure can be
summarized as follows:

\begin{enumerate}
\def\labelenumi{\arabic{enumi}.}
\tightlist
\item
  \textbf{Layer 1: Axiomatic Foundation (Appendix A -
  \texttt{MathLemmas})}: Established the fundamental, non-linear
  properties of integer arithmetic, including strict bounds on
  truncation loss.
\item
  \textbf{Layer 2: Financial Primitives (Appendix B, C -
  \texttt{AssetCalculations}, \texttt{Discounts})}: Built upon the
  axioms to prove the safety and correctness of isolated financial
  operations. Key properties included monotonicity and bounded
  round-trip safety.
\item
  \textbf{Layer 3: Composite Workflows (Appendix D, E, F, G, H -
  \texttt{Config}, \texttt{Deposit}, \texttt{Withdraw}, \texttt{Claim},
  \texttt{Distribution})}: Formalized the complete business logic for
  all user- and system-level interactions. This layer establishes
  end-to-end safety guarantees for every phase of the contract's
  lifecycle: from the atomicity and fund conservation of initial \emph{
  }deposits** (\texttt{Lemma\_RefundIsSafe}), through the accounting
  integrity of \textbf{withdrawals}, to the temporal correctness of
  post-sale \textbf{token claims} (vesting monotonicity) and the logical
  soundness of administrative \textbf{distributions}.
\item
  \textbf{Layer 4: Global State Machine (Appendix I -
  \texttt{Launchpad})}: Integrated all verified workflows into a global
  state machine, proving that the orchestration logic correctly and
  safely manages the system's overall state across its entire lifecycle.
\end{enumerate}

This hierarchical decomposition provides a robust and scalable
methodology for verifying mission-critical systems. The final safety
properties of the \texttt{Launchpad} contract are a logical and
inevitable consequence of the verified properties of its constituent
parts, culminating in a system with the highest possible degree of
formal assurance against a wide class of vulnerabilities, spanning
low-level financial arithmetic, complex business rule interactions, and
the complete state machine lifecycle.

\begin{center}\rule{0.5\linewidth}{0.5pt}\end{center}

\renewcommand\refname{References}
\bibliography{references.bib}

\thispagestyle{empty}

\end{document}